\documentclass[%
 reprint,
 superscriptaddress,
nofootinbib,
amsmath,amssymb,
aps,
pra,
floatfix,
]{revtex4-2}
\usepackage[utf8]{inputenc}
\usepackage{amsthm}

\usepackage[normalem]{ulem} 
\usepackage{natbib}		
\setcitestyle{square,numbers}
\usepackage[usenames,dvipsnames]{color}
\usepackage[margin=1.0in]{geometry}

\usepackage[breaklinks=true,%
            colorlinks=true,%
            linkcolor=blue,%
            urlcolor=blue,%
            citecolor=blue]{hyperref}
            
\usepackage{bbold} 

\newcommand{\bra}[1]{\ensuremath{\left\langle#1\right|}}
\newcommand{\ket}[1]{\ensuremath{\left|#1\right\rangle}}
\newcommand{\braket}[2]{\ensuremath{\left\langle#1\middle\vert#2\right\rangle}}

\usepackage{algpseudocode}
\usepackage{algorithm}

\usepackage{graphicx}
\usepackage{subcaption} 

\usepackage{tikz}
\usetikzlibrary{quantikz}

\begin{document}

\title{Advantages of multistage quantum walks over QAOA}

\author{Lasse Gerblich}

\affiliation{Department of Physics, University of Strathclyde, John Anderson Building, 107 Rotton Row East, Glasgow, G4~0NG, UK}

\author{Tamanna Dasanjh}
\email{current affiliation: University of Bristol}
\affiliation{Department of Physics 
Durham University, South Road, Durham, DH1~3LE, UK}

\author{Horatio Q. X. Wong}
\affiliation{Department of Physics
Durham University, South Road, Durham, DH1~3LE, UK}

\author{David Ross}
\affiliation{Department of Physics
Durham University, South Road, Durham, DH1~3LE, UK}

\author{Leonardo Novo}
\affiliation{International Iberian Nanotechnology Laboratory (INL), Av. Mestre José Veiga, 4715-330 Braga, Portugal}

\author{Nicholas Chancellor}
\email{nicholas.chancellor@gmail.com}
\affiliation{Department of Physics
Durham University, South Road, Durham, DH1~3LE, UK}\affiliation{School of Computing, Newcastle University, 1 Science Square, Newcastle upon Tyne NE4 5TG, United Kingdom}

\author{Viv Kendon}
\email{viv.kendon@strath.ac.uk}
\affiliation{Department of Physics
Durham University, South Road, Durham, DH1~3LE, UK}
\affiliation{Department of Physics, University of Strathclyde, John Anderson Building, 107 Rotton Row East, Glasgow, G4~0NG, UK}

\date{\today} 

\begin{abstract}
Methods to find the solution state for optimization problems encoded into Ising Hamiltonians are a very active area of current research.  In this work we compare the quantum approximate optimization algorithm (QAOA) with multi-stage quantum walks (MSQW).  Both can be used as variational quantum algorithms, where the control parameters are optimized classically. A fair comparison requires both quantum and classical resources to be assessed. Alternatively, parameters can be chosen heuristically, as we do in this work, providing a simpler setting for comparisons.   Using both numerical and analytical methods, we obtain evidence that MSQW outperforms QAOA, using equivalent resources. We also show numerically for random spin glass ground state problems that MSQW performs well even for few stages and heuristic parameters, with no classical optimization.
\end{abstract}

\maketitle

\section{Introduction}

While adiabatic quantum computing \cite{Farhi2000_AdiabaticQC,Kadowaki1998,Finnila1994} provides elegant intuition and proof for solving Hamiltonian problems, in practice, methods to obtain shorter run times are required to obtain a useful quantum advantage, especially in current quantum hardware with limited coherence time.
Diverse approaches have been proposed, including
diabatic quantum annealing \cite{Crosson2020} and
shortcuts to adiabaticity \cite{Cepaite2022}.
Continuous-time quantum walks with repeated short run times to amplify the probability of finding the ground state have also been shown to be effective
\cite{Callison2019,Moosavian2021,Banks2024continuoustime,Schulz2024guided}.
Hybrid quantum-classical methods can be leveraged to optimize the control parameters, including the quantum approximate optimization algorithm (QAOA) \cite{Farhi2014a,Farhi2014b,Hadfield17a,Banks2024rapidquantum},
which provides a route to implementation on gate-based architectures.

More recent detailed studies of the performance of QAOA and related algorithms include
\citeauthor{Marwaha2022} \cite{Marwaha2022} who derive bounds on QAOA performance,
\citeauthor{Atif2022suppressing} \cite{Atif2022suppressing} on suppressing unwanted fluctuations, and \citeauthor{Venuti2021optimal} \cite{Venuti2021optimal} who extend optimal control analysis to open systems relevant for experimental settings.
Experimental comparisons of QAOA and a variational quantum adiabatic algorithm (VQAA) have been carried out in Rydberg experiments 
\cite{Ebadi2022S}, in which they find VQAA performs better than QAOA in this setting.
Optimal quantum control theory has been used to prove the equivalence of QAOA to an optimal quantum annealing schedule in the limit of many stages \cite{Yang2017,Brady2021}.  

However, all methods for classically optimizing control parameters run the risk of doing all the work in the classical part of the algorithm, or using more prior knowledge of the problem than we should have access to, leading to methods that do not scale or generalize.
This makes it hard to correctly account for all the significant computational resources \cite{Schulz2024guided}.  
In this work, we carry out a comparison of multistage continuous-time quantum walks and QAOA, using specific (not necessarily variational) forms of quantum alternating operators ansatzes 
\cite{willsch2022gpu},
taking care to match like for like resources for the same problem.  
Guided by our analytical results in section \ref{sec:analytics}, which show that for many stages, both MSQW and QAOA approach optimal annealing schedules, we reduce the number of free parameters for our numerical comparisons of few stages in section \ref{sec:numerical}.
In both our numerical and analytical comparisons, we find that multi-stage quantum walks (MSQW) outperform QAOA for the same resources, as discussed in section \ref{sec:outlook} along with future research directions.
We give the relevant definitions and properties of adiabatic quantum optimization, diabatic quantum annealing, QAOA, and MSQW in section \ref{sec:background}, and describe our numerical methods in section \ref{sec:numerical_methods}. 
Further details of the calculations and numerical examples are given in appendices. 
\section{Background and definitions}\label{sec:background}
\subsection{Adiabatic quantum optimization}
Adiabatic quantum computation provides a conceptually simple strategy to solve combinatorial optimization problems based on the interpolation between a driver Hamiltonian $\hat{H}_d$ and a problem Hamiltonian $\hat{H}_P$
\begin{equation}\label{eq:Ham_adiabatic}
\hat{H}(t)= A(t) \hat{H}_{d}+ B(t) \hat{H}_{P}, 
\end{equation}
with the interpolation controlled by real, time-dependent parameters $A(t)$ and $B(t)$.
The problem Hamiltonian encodes a classical cost function with its ground state being the exact solution to a classical optimization problem. Typically, this is chosen as a classical Ising spin model mapped to qubits of the form
\begin{equation}\label{eq:Hprob}
\hat{H}_{P}= -\frac{1}{2}\sum_{a\neq b =0}^{n-1}J_{ab}\hat{Z}_a \hat{Z}_b-\sum_{b =0}^{n-1}h_b \hat{Z}_b,  
\end{equation} 
where $\hat{Z}_a$ is the Pauli-Z operator applied to the $a$th qubit, with $n$ being the number of qubits.
The real fields $h_a$ and coupling strengths $J_{ab}$ can be efficiently specified \cite{Choi2010adiabatic,Lucas2014ising} to encode solutions to many NP-complete problems in the ground state of $\hat{H}_p$.
The driver Hamiltonian is problem independent and allows the wave-function to explore the Hilbert space in superposition. We consider the standard choice for the driver Hamiltonian 
\begin{equation}\label{eq:Hdrive}
\hat{H}_{d}= - \sum_{j=0}^{n-1} \hat{X}_j, 
\end{equation}
where $\hat{X}_j$ are the Pauli-X operators acting on the $j$th qubit and couple bit-strings that differ by a single bit flip. The total Hamiltonian in equation~\eqref{eq:Ham_adiabatic} is then a time-dependent transverse field Ising model.

The adiabatic approach to prepare an exact solution to the problem is to start with the ground state of $\hat{H}_{d}$.  For our choice of $\hat{H}_{d}$, equation \eqref{eq:Hdrive}, this is the state 
$\ket{s}= \ket{+}^{\otimes n }$, with $\ket{+}= (\ket{0}+\ket{1})/\sqrt{2}$, and corresponds to the equal superposition of all $n$-bit strings.  
Then the Hamiltonian is slowly varied such that in the beginning of the procedure we have $A(t=0)=1$ and $B(t=0)=0$ while at the end $A(t=0)=0$ and $B(t=T)=1$.  This strategy is guaranteed to work with high success probability if the total evolution time $T$ is chosen proportional to $1/g_{\text{min}}^2$, with 
\begin{equation}
g_{\text{min}}=\text{min}_{0\leq t\leq T} (E_1(t)-E_0(t)) 
\end{equation} 
being the minimum gap between the ground state and first excited state throughout the evolution. This is the bottleneck of the running time of the algorithm -- for many problems $g_{min}$ is exponentially small which implies that a successful implementation of the algorithm would require exponentially long coherence times, which are experimentally impractical and may not provide any quantum advantage. 
The adiabatic model is conceptually and mathematically appealing, with an intuitive and mathematically accessible two-energy-level model. However, it is experimentally difficult to reach this regime, particularly with real hardware with limited coherence time. In fact, unless $NP \subseteq BQP$ (effectively the equivalent statement to $P=NP$, but when quantum devices are included as solvers), adiabatic run times would have to grow exponentially with problem size  \cite{Chancellor2020views,Callison2021}. 

\subsection{Diabatic Quantum Annealing}

In the past few years, different strategies inspired by the adiabatic approach have been put forward that are more suitable for implementations in near-term devices with short coherence times. 
In general, multiple techniques can be considered diabatic, including open systems exposed to thermal baths. A good overview is provided by \citeauthor{Crosson2020} \cite{Crosson2020}.  The techniques which are interesting for the purposes of this paper are annealing protocols which are based on rapid evolution, generally far from the adiabatic limit, where the adiabatic theorem does not even hold approximately, but where interactions with an external environment can be ignored. In this setting, the equations for problem and driver Hamiltonians are the same as in the adiabatic setting, equation~\eqref{eq:Ham_adiabatic}, but within a regime of shorter run times where the underlying physics is different. This is closer to the initial intuition behind quantum annealing \cite{Kadowaki1998}, which considered it as a quantum analogy of classical simulated annealing. It has been observed both numerically \cite{Callison2021} and experimentally \cite{Marshall2019pausing} that the intuition from the adiabatic setting \cite{Roland2002}, of slowing down at the point where the gap is the smallest, no longer necessarily holds in the diabtic setting. This justifies the need for different theoretical tools which are more suited to the diabatic regime. While the theory of diabatic annealing is less developed, partially because of the lack of simple gap-related arguments, energy related arguments appear to play a somewhat analogous role to the adiabatic argument \cite{Callison2021}.

In this work, we focus on two related diabatic strategies: QAOA and continuous-time quantum walks.  

\subsection{QAOA}
The quantum approximate optimization algorithm (QAOA) was proposed by \cite{Farhi2014a} \cite{Farhi2014a,Farhi2014b} as a heuristic optimization algorithm suitable for NISQ devices with a limited depth .  The algorithm can be viewed as a discretization of adiabatic quantum optimization as it consists of the alternate application of the two components of the Hamiltonian in equation \eqref{eq:Ham_adiabatic}, the driver Hamiltonian $\hat{H}_{d}$ from equation \eqref{eq:Hdrive} and the problem Hamiltonian $\hat{H}_P$ from \eqref{eq:Hprob}. 
A depth $p$ instance of a QAOA quantum circuit is defined by two sets of angles $\boldsymbol\alpha=(\alpha_1,...\alpha_p)$ and $\boldsymbol\beta=(\beta_1,...\beta_p)$ controlling the amount of each Hamiltonian to be applied at each stage $j\in\{1\dots p\}$, producing the output state
\begin{equation}\label{eq:QAOAdef}
\ket{\boldsymbol\alpha, \boldsymbol{\beta}}= \prod_{j=1}^p e^{- i \alpha_j\hat{H}_d}e^{- i \beta_j \hat{H}_{P}}\ket{s}, 
\end{equation}
with the initial state $\ket{s}= \ket{+}^{\otimes n }$, using units in which $\hbar=1$.   

The parameters $\boldsymbol{\alpha}$ and $\boldsymbol{\beta}$ are chosen to approximate the desired annealing schedule $A(t)$, $B(t)$.  In the VQA version of QAOA, this is done by using the quantum computer to calculate the energy expectation value with respect to the problem Hamiltonian $\hat{H}_P$
\begin{equation}
F_p(\boldsymbol{\alpha}, \boldsymbol{\beta}) = \bra{\boldsymbol{\alpha}, \boldsymbol{\beta}}\hat{H}_{P}\ket{\boldsymbol{\alpha}, \boldsymbol{\beta}},
\end{equation} 
followed by using a classical optimization algorithm, such as gradient descent, to update $\boldsymbol{\alpha}$ and $\boldsymbol{\beta}$.  This is repeated for as many rounds of optimization as required.  The aim is to reduce the energy expectation value $F_p(\boldsymbol{\alpha}, \boldsymbol{\beta})$, and thus output lower energy states closer to the solution state.  The energy expectation value $F_p(\boldsymbol{\alpha}, \boldsymbol{\beta})$ is used as a cost function to guide the classical optimization. 

Although well-optimized QAOA works well, finding the best choice of parameters can be costly, requiring many rounds of classical optimization, with corresponding calls to the quantum algorithm to update the expected energy $F_p(\boldsymbol{\alpha}, \boldsymbol{\beta})$.  The quantum cost of the optimization can thus dominate over the advantage from the final performance.  Numerical simulations have suggested that, for some problems, heuristic parameter choices based on discretization of linear adiabatic schedules may be preferable compared to trying to find the optimal parameters to run the algorithm~\cite{willsch2022gpu}, this protocol was termed approximate quantum annealing AQA. The protocols which we study numerically in section \ref{sec:numerical} can be understood as intances of AQA.

\subsection{Multi-stage quantum walks}

Continuous-time quantum walks have been widely used as a tool to construct quantum algorithms, providing exponential speed-ups for graph traversal problems \cite{Farhi1998_QW,Childs2003exponential}, and quadratic speed-ups for spatial search on graphs \cite{Childs2004}.
More recently, a few works have considered quantum walks as simple strategies for finding solutions to optimization problems. \citeauthor{Callison2019} \cite{Callison2019} investigated the performance of an algorithm based on a short time-evolution under a time-independent tranverse field Ising model of the form 
\begin{equation}\label{eq:QWopt}
\hat{H}_{QW}= \gamma \hat{H}_{d}+ \hat{H}_{P}. 
\end{equation} 
The dynamics can be seen as a quantum walk on the hypercube graph in Hilbert space, where the energy required for the walk to occupy a particular node is given by the problem Hamiltonian. It was shown \cite{Callison2019} that this simple approach, when applied to the problem of finding the lowest energy state of the Sherrington-Kirkpatrick spin glass Hamiltonian, is already able to outperform, on average, the bound given by Grover's algorithm which provides an exact solution in time $\tilde{O}(2^{n/2})$.
Classical algorithms can also beat Grover's algorithm \cite{Hartwig1984}, and \citeauthor{Montanaro2018} \cite{Montanaro2018} has combined both to provide a quantum advantage in a universal quantum computing setting. A related but distinct idea to multi-stage quantum walks which is potentially interesting for quantum circuit compilation is quantum walks on dynamic graphs \cite{Herrman2019dynamic}.


In this work, we generalize the quantum walk example presented in \citeauthor{Callison2021} \cite{Callison2021} which used two successive quantum walks with different hopping rates $\gamma_1$ and $\gamma_2$.
We define the multi-stage quantum walk (MSQW) optimization algorithm, in analogy with QAOA, as
\begin{equation}\label{eq:MSQW}
\ket{\boldsymbol{\gamma}, \boldsymbol{t}}= \prod_{j=1}^p e^{- i (\gamma_j \hat{H}_{d}+ \hat{H}_{p})t_j}\ket{s}, 
\end{equation}
where $\boldsymbol{\gamma} = (\gamma_1,\gamma_2\dots\gamma_p)$ is a set of hopping rates and $\boldsymbol{t}= (t_1,t_2\dots t_p)$ is a set of corresponding evolution times.

Using energetic arguments, \citeauthor{Callison2021} \cite{Callison2021} showed that the expected energy of the output state can only decrease with multiple stages of a quantum walk, provided the successive $\gamma_j$ are monotonically decreasing. Multi stage quantum walk algorithms have also been developed in \cite{Banks2024continuoustime,Schulz2024guided} using different parameterizations.  Although not the primary focus of their work, \citeauthor{Berwald2023Zeno} \cite{Berwald2023Zeno} demonstrated that, under very general circumstances, a MSQW with any number of stages can yield an optimal speedup on unstructured search. Effectively, the number of stages acts as an interpolation between adiabatic-like and quantum-walk-like dynamics, approaching the behavior of an adiabatic protocol as the number of stages becomes large. This is essentially a discretized version of the interpolation defined in \cite{Morley19interp}.

\section{Approximating annealing with MSQW and QAOA}\label{sec:analytics}

\citeauthor{Yang2017} \cite{Yang2017} and \citeauthor{Brady2021} \cite{Brady2021} used optimal control theory to show that QAOA as defined in equation \eqref{eq:QAOAdef} approximates an optimal quantum annealing schedule, with the best performance of QAOA obtained in the limit of a large number of stages $p\gg 1$ approaching a smooth anneal, apart from possible jumps at the start and end.
We carry out the same calculation for MSQW, which allows us to draw the same conclusions about MSQW for $p\gg 1$, i.e., that MSQW can also approximate an optimal quantum annealing schedule.
We then analytically derive the scaling with the number of approximation stages $p$ for how MSQW and QAOA approach a smooth quantum annealing schedule.  This allows us to directly compare the performance of MSQW and QAOA for many stages.

\subsection{MSQW as optimal protocols}\label{ssec:msqw_optimal}

We follow closely the analysis done by \citeauthor{Yang2017} \cite{Yang2017} and improved upon by \citeauthor{Brady2021} \cite{Brady2021} for Hamiltonians of the form in equation \eqref{eq:Ham_adiabatic} approximated by QAOA Hamiltonians as in equation \eqref{eq:QAOAdef}. 
Instead of the QAOA approximation, we use the MSQW approximation in equation \eqref{eq:MSQW}.
For ease of notation for the first part, where we derive the conditions for an optimal schedule, we rewrite equation \eqref{eq:Ham_adiabatic} as
\begin{equation}\label{eq:Ham_QW}
    \hat{H}(t) = \gamma(t)\hat{H}_d + \hat{H}_P,
\end{equation}
where $\gamma(t)$ can be any function, including piecewise linear QW segments in a MSQW.
This introduces a mild rescaling of time between the two forms of the Hamiltonians in equations \eqref{eq:Ham_adiabatic} and \eqref{eq:Ham_QW}, since $\gamma(t) \equiv A(t)/B(t)$.
Real hardware has maximum settings for the controls, which mean that the total energy of the Hamiltonian $\hat{H}(t)$ is capped in any realistic experimental setting.  Consequently, $A(t)$ and $B(t)$ are not fully independent parameters but must be kept within an overall bound.  Likewise, $\gamma(t)$ cannot be made arbitrarily large.  Adhering to this total energy constraint ensures the time rescaling between the two forms of the Hamiltonian does not affect the overall scaling comparisons we carry out here.
Further discussion of the relationship between the parameters $A(t)$, $B(t)$, $\gamma(t)$ and the consequences for the scaling of the run times can be found in \cite{Morley19interp}.

We can recast the optimisation problem as a control problem in the branch of mathematics called control theory. 
The aim is to find the control function $\gamma(t)$ that will minimize the quantity
\begin{equation}
    J = \langle\psi(t_f)|\hat{H}_P|\psi(t_f)\rangle
    \label{eq:minquant1}
\end{equation}
where $t_f$ is the total evolution time, and with the initial condition that we start in $|\psi(0)\rangle$. 
Note that the state $\psi(t_f)$ which minimises $J$ is the ground state of $\hat{H}_P$ and hence is the solution to the optimization problem.

We apply the constraint that the state evolves according to the Schr\"odinger equation 
\begin{equation}
    |\dot{\psi}\rangle = -i\hat{H}(t)|\psi\rangle,
    \label{eq:optev}
\end{equation}
where the overdot indicates a time derivative.
We can encode the constraint into the quantity we want to minimize by modifying equation \eqref{eq:minquant1}:
\begin{align}
\begin{split}
    J = &\langle\psi(t_f)|\hat{H}_P|\psi(t_f)\rangle\\
    &+ \int_0^{t_f} \/-\langle k|\dot{\psi} \rangle -i \langle k |\hat{H}(t) |\psi\rangle \mathrm{d}t +c.c. ,
\end{split}
    \label{eq:minquant2}
\end{align}
where $|k(t)\rangle$ is a Lagrange multiplier and $c.c.$ refers to the complex conjugate of the preceding term.

Variational calculus techniques (details in Appendix \ref{app:opt}) now give us some conditions on the Lagrange multiplier:
\begin{align}
    |\dot{k}(t)\rangle &= -i\hat{H}(t)|k(t)\rangle\nonumber\\
    |k(t_f)\rangle &= \hat{H}_P|\psi(t_f)\rangle
    \label{eq:FinLag}
\end{align}
Furthermore, we can define the \emph{control Hamiltonian} $\mathcal{H}$, which is not to be confused with any of the quantum mechanical Hamiltonians,
\begin{equation}
    \mathcal{H} = i\langle k|\hat{H}|\psi\rangle +c.c.
    \label{eq:contrHam}
\end{equation}
This is a conserved quantity, i.e., $\frac{\mathrm{d}\mathcal{H}}{\mathrm{d}t}=0$. 
Defining $\Phi_X(t)=i\langle k(t)|\hat{H}_X|\psi(t)\rangle+c.c.$, where $X$ identifies the Hamiltonian term(s) under consideration, we have
\begin{equation}
    \frac{\mathrm{d}\mathcal{H}}{\mathrm{d}t}=\Phi_d(t)\dot{\gamma}(t) = 0.
    \label{eq:TimeDev}
\end{equation}
From this, we can already deduce that we either require that $\dot{\gamma}(t)= 0$, which corresponds to a quantum walk segment, or that $\Phi_d(t)=0$. In the latter case \citeauthor{Brady2021} \cite{Brady2021} showed that this corresponds to quantum annealing for $\Phi_d(t)$ defined in the same way as done here.

We can rewrite equation \eqref{eq:contrHam} as 
\begin{equation}
    \mathcal{H} = \gamma(t)\Phi_d(t) + \Phi_P(t).
    \label{eq:ContrHam2}
\end{equation}
Making use of equations \eqref{eq:FinLag} we get that $\Phi_P(t_f)=0$ and therefore from equation \eqref{eq:ContrHam2}
\begin{equation}
    \mathcal{H} = \gamma(t_f)\Phi_d(t_f).
\end{equation}
A central result in control theory is Pontryagin's minimum principle \cite{Pontryagin1963}, which states that the optimal control function that minimises $J$ is the one that minimises the control Hamiltonian. From this, we can deduce that in order to minimise $\mathcal{H}$, a good approach is to make $\gamma(t_f)$ small.
Furthermore, without loss of generality we can define the driver Hamiltonian to have a zero ground state energy, so $\hat{H}_d|\psi(0)\rangle=0$. Thus, $\Phi_d(0)=0$, so at the starting time $\Phi_P(0)=\mathcal{H}$. Therefore, we do not need to restrict ourselves to small $\gamma_j$ in the beginning of the MSQW.

So far, we have deduced that $\Phi_P(t)$ goes from $\mathcal{H}$ in the beginning to $0$ in the end while $\Phi_d(t)$ starts off as $0$ and has some finite value in the end. We want this process to happen as quickly as possible, so ideally $\Phi_P(t)$ should decrease somewhat quickly. In order to achieve this, let us consider the time derivatives. Differentiating equation \eqref{eq:ContrHam2} with respect to time gives us
\begin{equation}
    \frac{\mathrm{d}\mathcal{H}}{\mathrm{d}t} = \dot{\gamma}(t)\Phi_d(t) + \gamma(t)\dot{\Phi}_d(t) + \Phi_P(t).
\end{equation}
Comparing this to equation \eqref{eq:TimeDev} we have
\begin{equation}
    \dot{\Phi}_P(t) = -\gamma(t)\dot{\Phi}_d(t).
\end{equation}
Hence, making $\gamma(t)$ big in the beginning while $\Phi_d(t)$ is still small allows us to quickly decrease $\Phi_P(t)$ while not making $\mathcal{H}$ too large.

In summary, a protocol which involves a QW step with large value for the hopping rate $\gamma$ in the beginning and a QW step with small $\gamma$ at the end satisfies the necessary requirements to be an optimal control for an optimization problem. However, as \citeauthor{Brady2021} \cite{Brady2021} have shown, the actual optimal control for QAOA has a smooth annealing schedule in the middle, so it is likely that the same is true here. More work would be needed to fully verify this for MSQW.

\subsection{Approximation of quantum annealing with QAOA and MSQW}\label{ssec:qa_approx}

We now consider in more detail how QAOA and MSQW approximate quantum annealing as the number of stages is varied.  The goal is to directly compare QAOA and MSQW in the same setting using the same quantum and classical resources.  The calculation is outlined here, with more details given in Appendix \ref{app:approx}.

Our analysis proceeds by expanding unitary time evolution operators in their Dyson series.  To do this, we need the time they are applied for to be small enough for the series to converge.
We therefore begin by dividing the total run time $t_p$ into $p$ subintervals $(t_0,t_1),\dots(t_{p-1},t_p)$ and expanding the quantum annealing evolution $\hat{U}_{QA}^{(j)}$ for a single subinterval $j$ into its Dyson series:
\begin{align}
\begin{split}
    \hat{U}_{QA}^{(j)} &= \mathcal{T}\left[e^{-i\int_{t_{j-1}}^{t_j}\hat{H}_{QA}(t) \,\mathrm{d}t}\right]\\
&= \mathbb{1} -i\int_{t_{j-1}}^{t_j}\hat{H}_{QA}(t) \,\mathrm{d}t\\
&\quad- \int_{t_{j-1}}^{t_j}\int_{t_{j-1}}^t\hat{H}_{QA}(t)\hat{H}_{QA}(t') \,\mathrm{d}t'\,\mathrm{d}t+\dots\\
&= \mathbb{1} -i\tilde{H}_j(t_j-t_{j-1}) -\frac{1}{2}\tilde{H}_j^2(t_j-t_{j-1})^2 + \dots \\
&\quad+\mathcal{O}\left(H_{max} \dot{H}_{max}(t_j-t_{j-1})^3\right),
\end{split}
\label{eq:Dyson}
\end{align}
where we defined 
\begin{equation}
\tilde{H}_j =\frac{1}{t_j-t_{j-1}}\int_{t_{j-1}}^{t_j}\hat{H}(t)\,\mathrm{d}t = \alpha_j\hat{H}_d+\beta_j\hat{H}_P    
\end{equation}
as well as 
\begin{align}
    H_{max}&=\max\limits_{t\in(t_0,t_f)}\|\hat{H}(t)\|\\
    \dot{H}_{max}&=\max\limits_{t\in(t_0,t_f)}\|\frac{d}{dt}\hat{H}(t)\|.
    \end{align}
Throughout this section, we assume the time steps $\Delta t_j =t_j-t_{j-1}$ are chosen such that $H_{max}\Delta t_j\ll 1$ and $\dot{H}_{max}(\Delta t_j)^2\ll 1$ for the expansion to be a good approximation of the true dynamics happening at each time-step. The second order expansion of equation \eqref{eq:Dyson} is the baseline against which the QAOA and MSQW approximations are to be compared.

Turning to the Dyson expansion for the QAOA evolution operator derived from equation \eqref{eq:QAOAdef} for the segment $(t_{j-1},t_j)$, using the same parameters $\alpha_j$ and $\beta_j$, gives
\begin{align}
\begin{split}
    \hat{U}_{QAOA}^{(j)} &= e^{- i \alpha_j\hat{H}_d(t_j-t_{j-1})}e^{- i \beta_j \hat{H}_{P}(t_j-t_{j-1})}\\
    &= \mathbb{1} -i\tilde{H}_j(t_j-t_{j-1}) -\frac{1}{2}\tilde{H}_j^2(t_j-t_{j-1})^2\\
    &\quad-\frac{1}{2}\alpha_j\beta_j \left[\hat{H}_d,\hat{H}_P\right](t_j-t_{j-1})^2\\
    &\quad+ \mathcal{O}(H_{max}^3(t_j-t_{j-1})^3)
\end{split}
\label{eq:QAOASegment}
\end{align}
 
Comparing equation~\eqref{eq:Dyson} and equation~\eqref{eq:QAOASegment}, the difference between the QA and QAOA evolution operators is thus
\begin{align}
\begin{split}
    &\|\hat{U}_{QA}^{(j)}-\hat{U}_{QAOA}^{(j)}\|= \frac{1}{2}\left\|\left[\hat{H}_d,\hat{H}_P\right]\right\|\alpha_j\beta_j(t_j-t_{j-1})^2\\
    &\quad+\mathcal{O}\left(H_{max}^3(t_j-t_{j-1})^3 + H_{max} \dot{H}_{max}(t_j-t_{j-1})^3\right)
\end{split}
\label{eq:QAvsQAOA}
\end{align}
We have chosen to use the operator norm here since 
\begin{align*}
\begin{split}
\|\psi_{QA}(t_j)-\psi_{QAOA}(t_j)\| &\leq \|\hat{U}_{QA}^{(j)}-\hat{U}_{QAOA}^{(j)}\| \|\psi(t_{j-1})\|,
\end{split}
\end{align*}
and $\|\psi(t_{j-1})\|=1$,
so it directly gives us an upper bound for the difference between the states we end up in after letting the system evolve via quantum annealing versus QAOA.

If we make all subintervals the same length, i.e., $(t_j-t_{j-1})=\frac{(t_f-t_0)}{p}$ for all $j$, then we can estimate the error in the total time evolution by adding up the errors for each subinterval, giving
\begin{align*}
&\|\hat{U}_{QA}(t_0,t_f) - \hat{U}_{QAOA}(t_0,t_f)\|\\ &\leq \frac{1}{2}\left\|\left[\hat{H}_d,\hat{H}_P\right]\right\|\sum_{j=1}^p\alpha_j\beta_j \left(\frac{t_f-t_0}{p}\right)^2\\
&\quad+\mathcal{O}\left(H_{max}^3\frac{(t_f-t_0)^3}{p^2} + H_{max} \dot{H}_{max}\frac{(t_f-t_0)^3}{p^2}\right)  
\end{align*}
The sum term scales linearly with $p$, so overall the QAOA algorithm approaches the annealing schedule as $\frac{1}{p}$, since, especially for large $p$, this will dominate over the $\frac{1}{p^2}$-terms. 
This means that the more subintervals we divide the evolution into, the better the approximation, and therefore the more likely we are to end up in the desired ground state.


In order to improve this result we can use the second-order Suzuki formula \cite{Childs2021_TrotterError} to approximate each subinterval:
\begin{align*}
    \hat{U}_{QAOA^\ast}^{(j)}&=e^{-i\alpha_j\hat{H}_d\frac{(t_j-t_{j-1})}{2}}e^{-i\beta_j\hat{H}_P(t_j-t_{j-1})}\\
    &\quad \times e^{-i\alpha_j\hat{H}_d\frac{(t_j-t_{j-1})}{2}}\\
    &= \mathbb{1} - i\tilde{H}(t_j-t_{j-1}) - \frac{1}{2}\tilde{H}^2(t_j-t_{j-1})^2\\
    &\quad + \mathcal{O}\left(H_{max}^3(t_j-t_{j-1})^3\right).
\end{align*}
We note that a sequential application of unitaries of the form above would still result in a QAOA-like quantum circuit. This results in a cumulative error for the whole anneal of
\begin{align}
\begin{split}
    &\|\hat{U}_{QA}(t_0,t_f)-\hat{U}_{QAOA}(t_0,t_f)\| \\
    &= \mathcal{O}\left(H_{max}^3\frac{(t_f-t_0)^3}{p^2}+H_{max}\dot{H}_{max}\frac{(t_f-t_0)^3}{p^2}\right),
\end{split}
\label{eq:ModQAOA}
\end{align}
so this modified QAOA approaches the annealing schedule as $\frac{1}{p^2}$, instead of $\frac{1}{p}.$
The two error terms appearing in the previous expression can be easily understood. The first is a Trotterization error, coming from the fact that QAOA circuits cannot directly apply weighted sums of the two non-commuting Hamiltonians $\hat{H}_P$ and $\hat{H}_d$. The second term is proportional to the time-derivative of $\hat{H}(t)$, and comes from the fact that we are neglecting the time-dependence of the Hamiltonian at each time-step by choosing constant parameters $\alpha_j$ and $\beta_j$ to parametrize the QAOA circuit. The MSQW approach does not suffer from the first type of error, resulting in a better simulation, as we explicitly verify in what follows. 

To analyse the performance of MSQW as a way to simulate quantum annealing, we choose to rewrite the quantum walk Hamiltonian in Eq. \eqref{eq:QWopt} as
\begin{equation}
    \hat{H}_{QW} = \alpha\hat{H}_d+\beta\hat{H}_P,
\end{equation}
where the notation using $\alpha,\beta$ is deliberately similar to the QAOA parameters $\boldsymbol{\alpha},\boldsymbol\beta$.
As already discussed, this is a mild rescaling of the Hamiltonian in equation \eqref{eq:QWopt}. 
To recover the quantum walk dynamics as defined in equation \eqref{eq:Ham_QW}, we rescale the time intervals for each stage, i.e., we let the system evolve for time $\beta t$ instead of $t$.  However, for comparisons in which the total energy of the Hamiltonian is capped, the $\alpha$, $\beta$ form of the equations is easier to work with.
Hence, the evolution operator for a single stage $j$ of the multi-stage quantum walk can be written
\begin{equation*}
    \hat{U}_{QW}^{(j)} = e^{-i(\alpha_j\hat{H}_d+\beta_j\hat{H}_P)(t_j-t_{j-1})}.
\end{equation*}
Expanding this gives
\begin{equation}
    \hat{U}_{QW} = \mathbb{1} -i\tilde{H}_j(t_j-t_{j-1}) -\frac{1}{2}\tilde{H}_j^2(t_j-t_{j-1})^2 + \dots
\end{equation}
Hence, we have
\begin{align*}
    &\|\hat{U}_{QA}(t_0,t_f)-\hat{U}_{QW}(t_0,t_f)\|\\
    &=\mathcal{O}\left(H_{max}\dot{H}_{max}\frac{(t_f-t_0)^3}{p^2}\right)
\end{align*}
From this we deduce that MSQW for a given number of stages performs better than the QAOA algorithm, since it does not have the extra term in equation \eqref{eq:QAvsQAOA} or equation \eqref{eq:ModQAOA}. Furthermore, like the modified QAOA, MSQW approaches quantum annealing faster than the first order Lie-Trotter expansion, as $\frac{1}{p^2}$ instead of $\frac{1}{p}$. \\
Therefore, we would expect an algorithm consisting of multiple quantum walk stages to solve the optimization problem with higher accuracy than the QAOA algorithm, for the same number of stages. 

Lastly, note the dependence on the number of qubits. Both $\|\hat{H}_d\|$ and $\|\hat{H}_P\|$ correspond to the largest eigenvalue of $\hat{H}_d$ and $\hat{H}_P$, respectively, the former of which always increases with the number of qubits.  Hence, for higher dimensions we would need to perform more stages than for lower dimensions in order to make the upper bound on the error similarly tight.

\section{Numerical methods}\label{sec:numerical_methods}

The numerical simulations to obtain our results in sections \ref{sec:numerical} and \ref{sec:reduced} were run on desktop workstations and on HPC facilities based at Durham University (``Hamilton''), and at the University of Strathclyde (``wee-ArCHIE'').
Our code was written in Python3 \cite{Python},

NumPy \cite{Numpy} and pandas \cite{Pandas} packages were used for data processing. Plotting was done using matplotlib \cite{matplotlib}.
Where relevant, figures have error bars, although the size of these is often too small to be seen.  For data sets of $10^4$ instances, the expected error in the averages is around 1\%.

We used sizes $5 \leq n \leq 10$ Sherrington-Kirpatrick (SK) Ising spin glass data sets of $10^4$ instances per size, from data archive: \cite{Callison2019_dataset}.
These SK spin-glass models \cite{Kirkpatrick1975_solvable} are characterized by a Hamiltonian of the form of equation \eqref{eq:Hprob}with couplings  $J_{jk}$ and fields $h_j$ drawn from a normal distribution with standard deviation $\omega_{SK}$, an arbitrary energy unit set to $\omega_{SK}=1$.

\section{Numerical comparisons}\label{sec:numerical}

The analytical scaling results from section \ref{ssec:qa_approx} strongly suggest that MSQW simulate quantum annealing schedules better than QAOA for the same resources, if we can apply these protocols for many stages.  However, the analytical results strictly only hold for $p\gg 1$, and, for the practical regimes of a few stages $1\le p\le 5$, it needs to be checked by other methods whether significant differences in performance can be observed.  We have therefore carried out detailed numerical comparisons for small values of $p$, starting with $p=1$, and continuing with $p=2$ and $p=5$.

Implementing both halves of the Hamiltonian together as in MSQW is a more complex operation than applying the separate pieces in turn, but current quantum annealers have this capability already in their normal operation. Gate model hardware can in principle apply more QAOA stages to compensate for the lack of simultaneous implementation, but this comes with a cost of a larger circuit depth, which impacts on performance in NISQ era machines.

Classically optimizing the parameters for a small number of stages is also a significant cost for running the algorithm, since each step in the optimization requires a quantum run to evaluate the cost function.  The optimization costs can easily become a significant or dominant part of the total resources, c.f., \cite{Schulz2024guided}. 
We therefore also evaluated ways to reduce and simplify the required parameters for larger values of $p$, while still obtaining good performance.

\subsection{Single-stage QW vs QAOA}

While single-stage QAOA is not expected to be efficiently simulable on classical computers  \cite{farhi2016quantum, DiezValle2023}, classical algorithms seem to be able to outperform it in this shallow depth regime \cite{hastings2019classical}. A single stage quantum walk is fundamentally different since the application of both the driver and problem Hamiltonian at the same time allows feedback between the two halves, whereas a single stage of QAOA does not. 

For a single-stage, both the quantum walk and QAOA are controlled by two parameters $(\gamma,t)$ or $(\alpha,\beta)$ respectively, making it practical to visualize the variation in the expectation value of the problem Hamiltonian $\hat{H}_P$ with the final state, i.e., the energy of the final state,
\begin{equation}
F^{QAOA}_P(\alpha,\beta) = \bra{\alpha, \beta}\hat{H}_{P}\ket{\alpha, \beta}
\end{equation}
and 
\begin{equation}
F^{QW}_P(\gamma, t) = \bra{\gamma, t}\hat{H}_{P}\ket{\gamma, t}.
\end{equation}
This provides a straightforward figure of merit for the algorithms, a lower final energy w.r.t.~the problem Hamiltonian quantifies how well the algorithm finds low energy states.
We are also interested in comparing the probability of obtaining the exact solution of the problem, i.e., the ground state of $H_P$ which we denote as the state $\ket{z*}$. The success probabilities for each algorithm to find this ground state are thus defined as
\begin{equation}
P^{QAOA}_P(\alpha, \beta)
 = |\braket{\alpha, \beta}{z*}|^2
\end{equation}
and 
\begin{equation}
P^{QW}_P(\gamma, t)
 = |\braket{\gamma, t}{z*}|^2, 
\end{equation}
for single-stage QAOA and single-stage QW, respectively.
%
\begin{figure*}[htp!]
        \centering
        \begin{subfigure}[b]{0.475\textwidth}
            \centering
            \includegraphics[width=\textwidth]{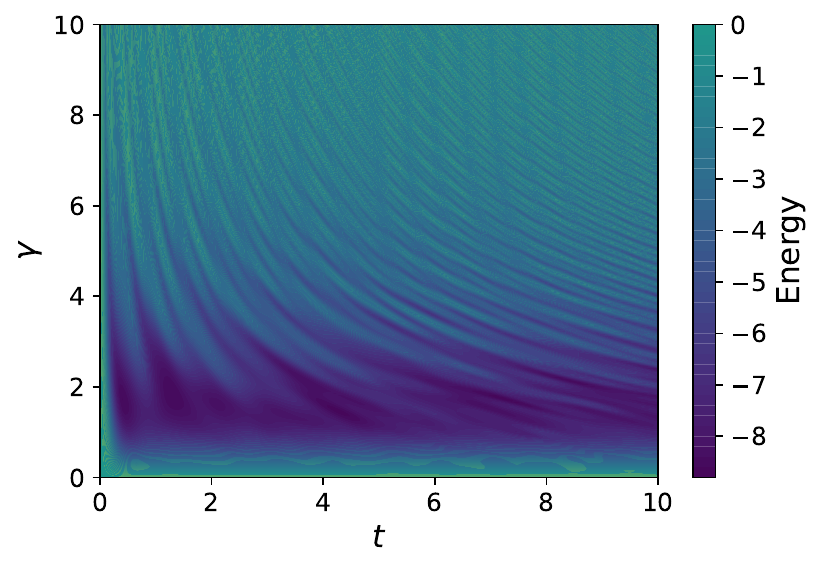}
            \caption[QW energies]%
            {QW energies}
            \label{fig:1_stage_QW_E}
        \end{subfigure}
        \hfill
        \begin{subfigure}[b]{0.475\textwidth}  
            \centering 
            \includegraphics[width=\textwidth]{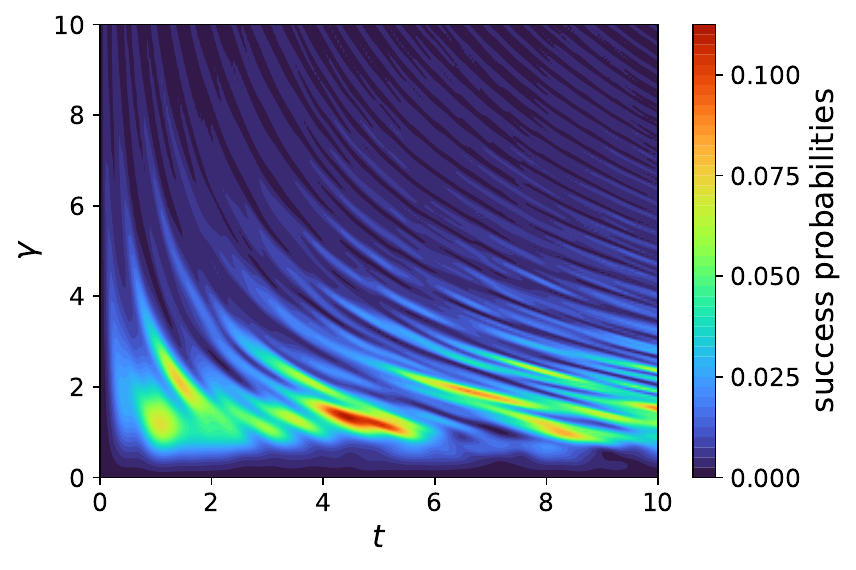}
            \caption[QW probabilities]%
            {QW probabilities}   
            \label{fig:1_stage_QW_prob}
        \end{subfigure}
        \vskip\baselineskip
        \begin{subfigure}[b]{0.475\textwidth} 
            \centering 
            \includegraphics[width=\textwidth]{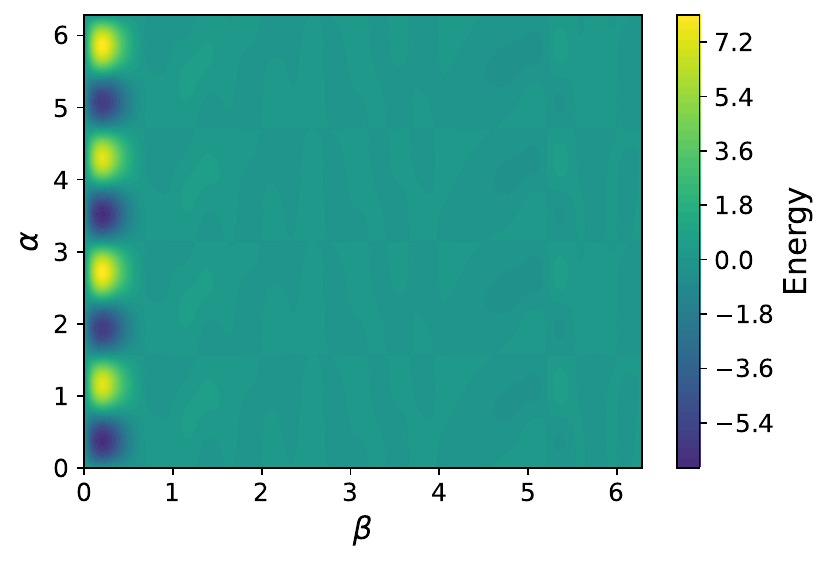}
            \caption[QAOA energies]%
            {QAOA energies}    
            \label{fig:1_stage_QAOA_E}
        \end{subfigure}
        \hfill
        \begin{subfigure}[b]{0.475\textwidth}  
            \centering 
            \includegraphics[width=\textwidth]{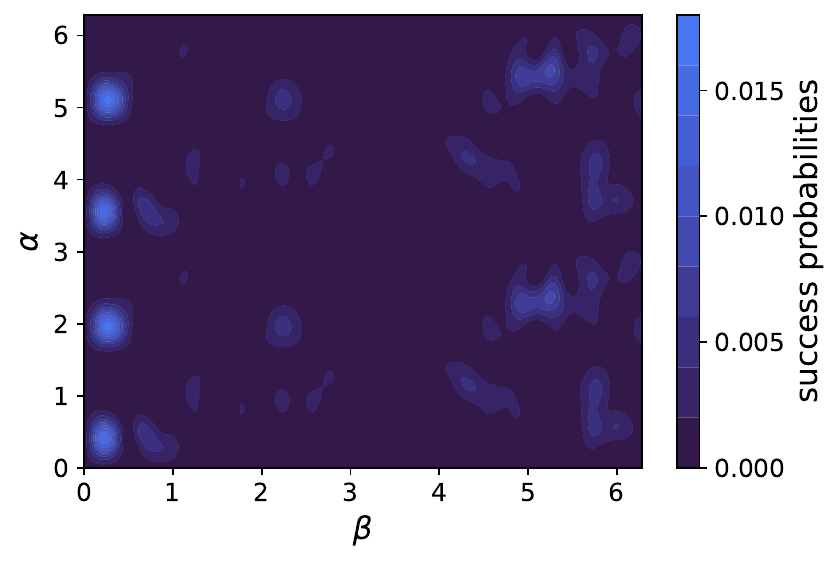}
            \caption[QAOA probabilities]%
            {QAOA probabilities}    
            \label{fig:1_stage_QAOA_prob}
        \end{subfigure}
        \caption[single_stage]
        {Quantum walk and QAOA average energies and success probabilities for a single stage. For quantum walk, $\gamma$ and $t$ are used to parameterise the single stage protocol. For QAOA,  $\alpha$ is the time for which the driver is active, and $\beta$ is the time for which the problem Hamiltonian is active. All plots are for for the 10 qubit SK spin glass instance with ID: aaaufeflfwqdwhthcrcnynopihzciv. Energy and success probability plots share the same color scale to emphasise the performance difference between the two protocols.} 
        \label{fig:single_stage}
\end{figure*}
We plot these two quantities for the quantum walk and QAOA in Fig.~\ref{fig:single_stage}, for a range of the relevant parameters.  Instead of optimizing $\gamma$, $t$ and $\alpha$, $\beta$, we numerically evaluated the whole range discretized over $20$ 
divisions in each direction. These divisions were linearly spaced in the range $0\le \gamma \le 4$ and $0\le t \le 6$ for quantum walks, and $0\le \alpha \le \pi/2$ and $0\le \beta \le \pi/2$ for QAOA. The plots then show where the optimal values lie, and the shape of the landscape around them.
Figure \ref{fig:single_stage} clearly shows that, for optimal parameters, QW finds lower energy states with higher probability than single stage QAOA.  
For a fixed $\gamma$ in the optimal region, ($1 \lesssim\gamma\lesssim 2$), we see an oscillatory behavior of both the expected energy $F_1^{QW}(\gamma, t)$ as well as of the success probability. 
Single stage QAOA has a very different energy landscape, with a few localized regions of parameters that leads to the best performance. Overall, we see that the quantum walk performs significantly better, reaching a global minimum for the expected energy of $\approx -8.76$, while QAOA reaches only $\approx -6.70$. Moreover, for a wide range of QW parameters, this algorithm can also find the ground state of $\hat{H}_P$ with significantly higher success probabilities than the best possible success probability of single-stage QAOA. 
Since we are doing a brute force scan of the parameter space in our simulations as our method of finding the optimal parameters, the cost of the optimization rounds to obtain these are a separate consideration we do not compare here.

\begin{figure*}[htp!]
        \centering
        \begin{subfigure}[b]{0.475\textwidth}
            \centering
            \includegraphics[width=\textwidth]{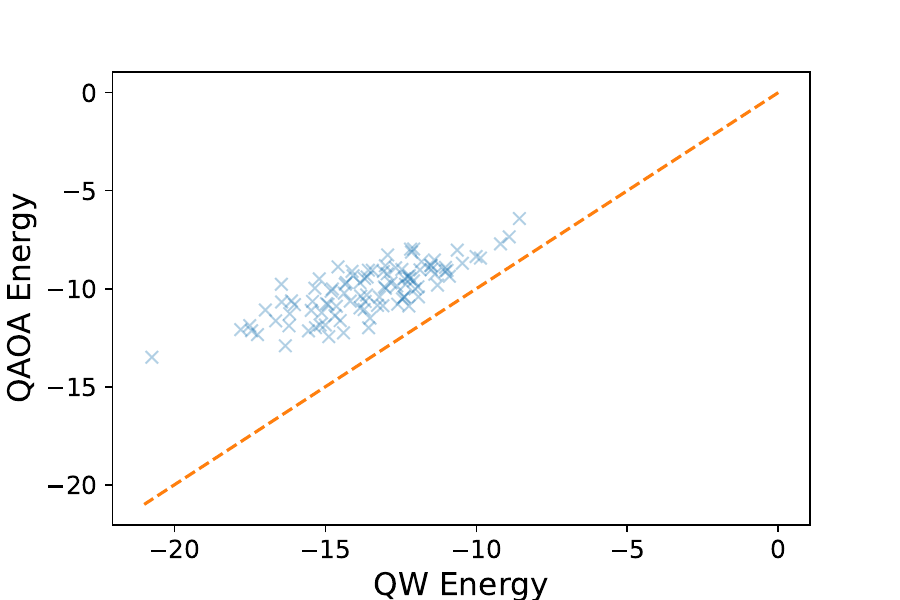}
            \caption[QW energies]%
            {Energies}
            \label{fig:1_stage_E_many}
        \end{subfigure}
        \hfill
        \begin{subfigure}[b]{0.475\textwidth}  
            \centering 
            \includegraphics[width=\textwidth]{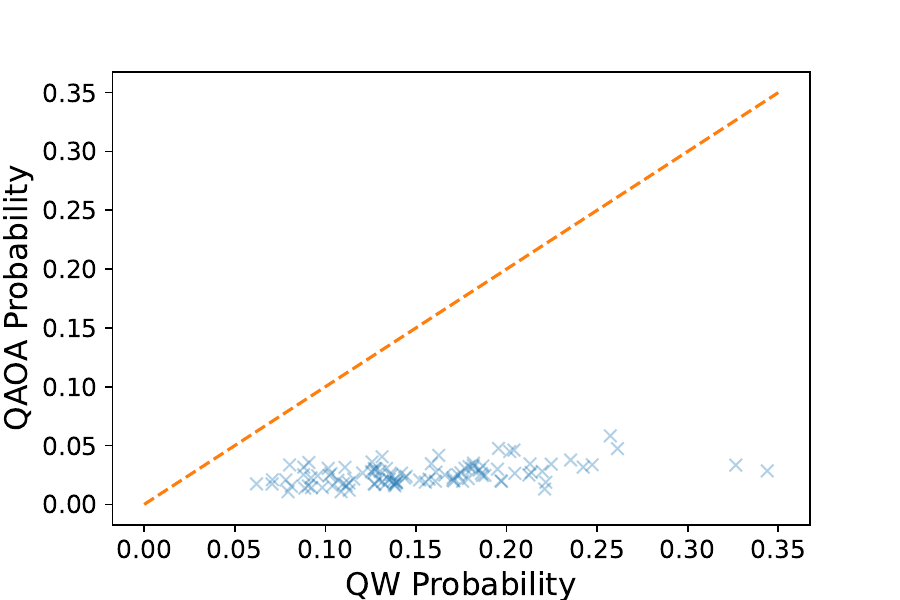}
            \caption[scatter probabilities]%
            {Probabilities}   
            \label{fig:1_stage_prob_many}
        \end{subfigure}
        \caption[single_stage]
        {Comparison of final problem Hamiltonian energy expectations (a) and success probabilities (b) for single stage QW and QAOA. Symbols represent the first $100$ ten qubit problems in the dataset from \cite{Callison2019_dataset}. Optimal parameters were found using a $20$ by $20$ grid search with an evenly-spaced grid with $0 \le \gamma \le 4$ and $0 \le t \le 6 $ for QW and $0 \le \alpha \le \pi/2 $ and $0 \le \beta \le \pi/2$ for QAOA. Dashed lines are a guide to the eye showing what equal performance would look like.}
        \label{fig:single_stage_many}
\end{figure*}

Although the plotted data in Fig.~\ref{fig:single_stage} is for a single instance of a random spin glass, we observed the trend that the quantum walk outperforms QAOA on multiple instances. By numerically comparing single stage versions of MSQW and QAOA for single instances 
from the SK dataset \cite{Callison2019_dataset}, we find that for 
the first $100$ Hamiltonians, an optimally parameterised single stage QW always outperforms an optimally parameterised single stage QAOA both in terms of average energy and success probability. The results are shown in scatter plots in Fig.~\ref{fig:single_stage_many}. This confirms that, even with classical optimization of the QAOA parameters, single stage QAOA is not as effective an algorithm for solving hard optimization problems.

\subsection{Two-stage MSQW vs QAOA}\label{sec:two-stage}

%
\begin{figure*}[htp!]
        \centering
        \begin{subfigure}[b]{0.475\textwidth}
            \centering
            \includegraphics[width=\textwidth]{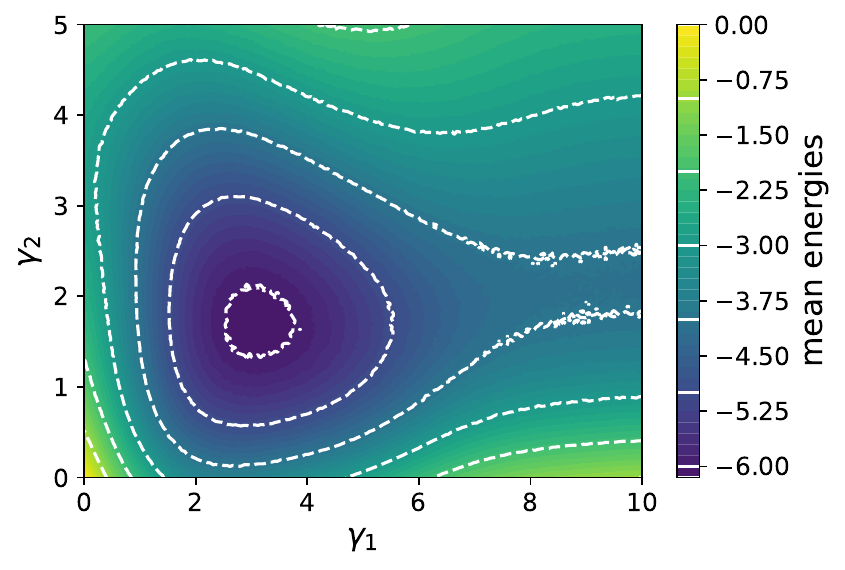}
            \caption[QW energies]%
            {QW energies}
            \label{fig:2_stage_QW_E}
        \end{subfigure}
        \hfill
        \begin{subfigure}[b]{0.475\textwidth}  
            \centering 
            \includegraphics[width=\textwidth]{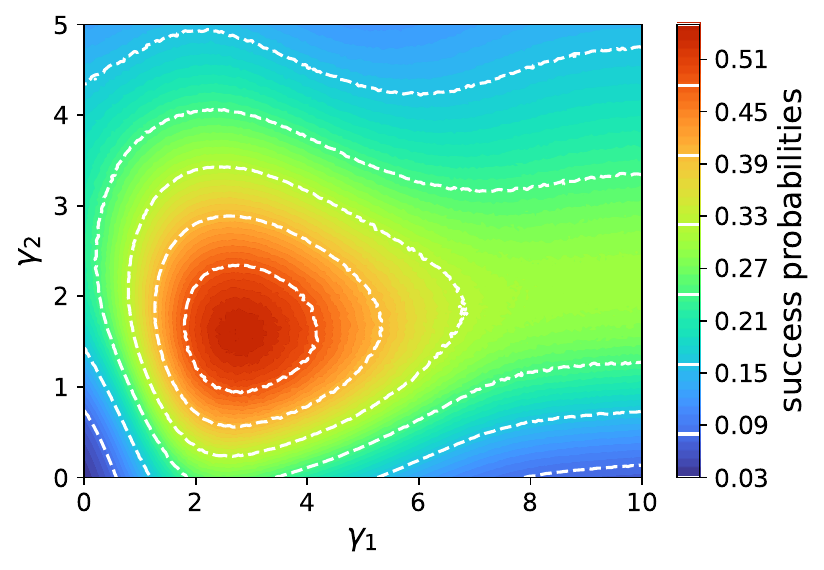}
            \caption[QW probabilities]%
            {QW probabilities}   
            \label{fig:2_stage_QW_prob}
        \end{subfigure}
        \vskip\baselineskip
        \begin{subfigure}[b]{0.475\textwidth} 
            \centering 
            \includegraphics[width=\textwidth]{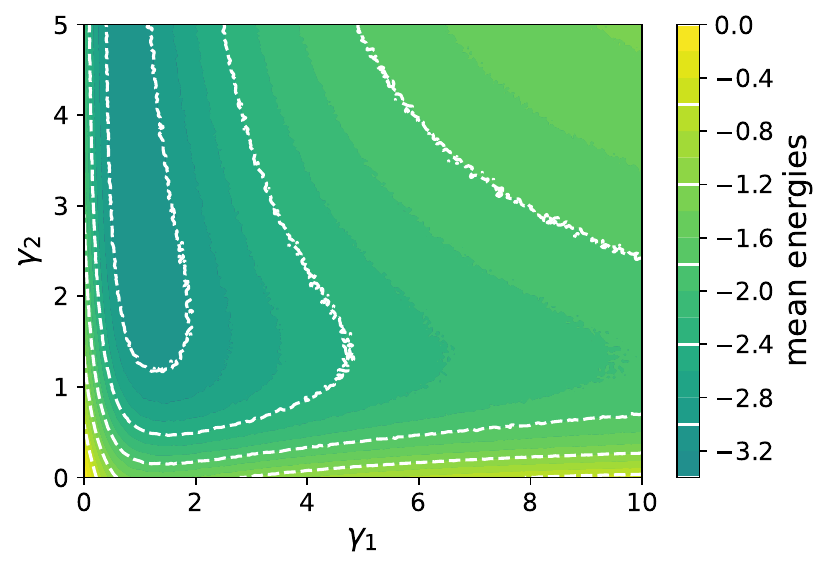}
            \caption[QAOA energies]%
            {QAOA energies}    
            \label{fig:2_stage_QAOA_E}
        \end{subfigure}
        \hfill
        \begin{subfigure}[b]{0.475\textwidth}  
            \centering 
            \includegraphics[width=\textwidth]{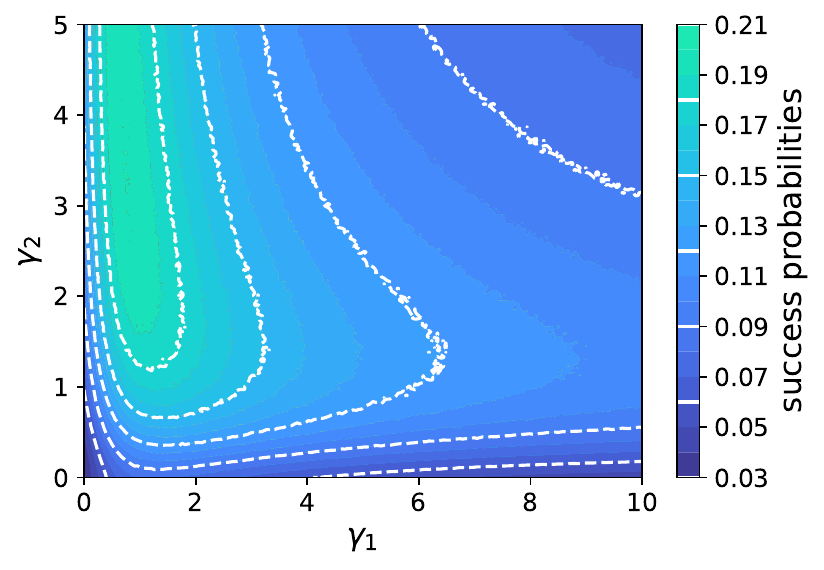}
            \caption[QAOA probabilities]%
            {QAOA probabilities}    
            \label{fig:2_stage_QAOA_prob}
        \end{subfigure}
        \caption[single_stage]
        {Quantum walk and QAOA average energies and success probabilities for two stages. For quantum walk, $\gamma$ and $t$ are used to parameterise the single stage protocol, which induce an analogous parametrization for QAOA in terms of these parameters (see equation~\eqref{eq:param_QAOA}). After time-averaging,  both protocols are parameterised in terms of $\gamma_1$ and $\gamma_2$ as described in the main text, and averaged over 2,000 samples with independent random runtimes for each stage in the range $0.1-0.5$. All plots are for for the $5$ qubit SK spin glass instance with ID: aaavmaiqiolnplcovmzxjazkyvyayz. Energy and success probability plots share the same color scale to emphasise the performance difference between the two protocols. 

        }
        \label{fig:two_stage}
\end{figure*}

An improvement of the performance of both the quantum walk and QAOA approach can be obtained by considering additional stages. Here, we analyse the two-stage case, focusing on strategies that allow us to reduce the number of free parameters. Two-stage quantum walks and QAOA each have four free parameters (within an overall energy cap), which makes visualising the dependence more challenging.  For quantum walks, we use the result in \cite{Callison2019} that quantum walks quickly (in time $\mathcal{O}(n)$ where $n$ is the number of spins) reach a regime where the time averaged success probability is stable.  This allows us to randomly sample over short times for half of the parameters, i.e., $t_1$ and $t_2$, leaving two free parameters $\gamma_1$ and $\gamma_2$, which can be visualised easily.
As we have seen in Fig.~\ref{fig:single_stage}, both the expected energy and success probability exhibit an oscillatory behavior. Hence, averaging over time not only allows us to remove one free parameter but also to smooth the energy landscape to better extract the overall behavior of the quantum walk.
Of course, in practice, we will need some repeats of the algorithm to achieve this performance, but repeats are required for any non-deterministic algorithm to ensure a high probability of success.

We plot the energy in Fig.~\ref{fig:2_stage_QW_E} and probability landscape in Fig.~\ref{fig:2_stage_QW_prob} for the time-averaged two-stage quantum walk, for a 5-qubit instance of the SK model.  The counterpart of these plots for two-stage QAOA is presented in Figs.~\ref{fig:2_stage_QAOA_E} and \ref{fig:2_stage_QAOA_prob}. In order to do a fair comparison, we parametrize the QAOA algorithm in a way that resembles the QW parametrization. Namely, we choose the parameters defining the $j$th stage of QAOA as
\begin{align}
    \alpha_j&= \frac{\gamma_j t_j}{1+\gamma_j} \nonumber\\
    \beta_j&= \frac{t_j}{1+\gamma_j}.  
    \label{eq:param_QAOA}
\end{align}
This parametrization ensures that the total evolution time $\alpha_j+\beta_j=t_j$ (and that the energy of the full Hamiltonian is capped, as discussed in section \ref{sec:analytics}). Moreover, it guarantees the same relative strength between the driver and problem Hamiltonian as in the $j$th stage of the QW, since $\alpha_j/\beta_j=\gamma_j$.

Figures \ref{fig:1_stage_QAOA_E} and \ref{fig:1_stage_QAOA_prob} show numerically that the performance for a single QAOA stage is especially poor if the problem Hamiltonian is allowed to act for too long. The reason for this is that the energies are encoded in the phase. It is only possible to unambiguously distinguish phase up to an additive factor of $2 \pi m$, where $m$ is an integer. A phase rotation of more than $\pi$ is therefore indistinguishable from a smaller phase rotation in the opposite direction (note that both positive and negative phase rotations are possible). Therefore, for any state where $|\beta\Delta E|>\pi$ where $\Delta E$ is the difference between the ground state and a given excited state, the energy will not be faithfully represented. 

More concretely, for $0<\alpha<\pi/4$, the off diagonal elements in this part of the unitary will be negative multiples of the imaginary unit and therefore positive or negative interference will be determined by the relative sign of the imaginary parts of the two diagonal elements.  For $2 \pi>\beta\Delta E>\pi$ the sign will be opposite, leading to overall constructive interference despite $\Delta E$ being positive and the transition therefore being to a higher energy state. Conversely, if  $-2 \pi<\beta\Delta E<-\pi$, the imaginary components will have the same sign, implying destructive interference, when the interference should be constructive given that $\Delta E<0$. 
To avoid these issues, we have limited the total runtime to $\le 0.5$, and have averaged over runtimes between $0.1$ and $0.5$, which we have empirically found to work well. 

As seen in Figs.~\ref{fig:1_stage_QW_E} and \ref{fig:1_stage_QW_prob}, quantum walk does not have this issue, since the driver and problem Hamiltonians are simultaneously active.  Nonetheless, we have applied the same runtime limits to our quantum walk simulations in the interest of fair comparison. We tested the validity of this approach by averaging over longer runtimes for quantum walk, and as is expected from the fast equilibration times observed in \cite{Callison2019}, there is essentially no difference between averaging over short and much longer runtimes.

This simple strategy to reduce the number of free parameters works well for the two-stage quantum walk, leading to a smooth optimization landscape, revealing a large region of high success probabilities and low values of the cost function.  This is in contrast with QAOA, which exhibits significantly higher values of the cost function and lower values of success probabilities. This suggests that two-stage QAOA requires a more careful optimization over all four parameters to achieve a good performance.

\section{Fewer parameters per stage}\label{sec:reduced}

\begin{figure*}[htp!]
        \centering
        \begin{subfigure}[b]{0.475\textwidth}
            \centering
            \includegraphics[width=\textwidth]{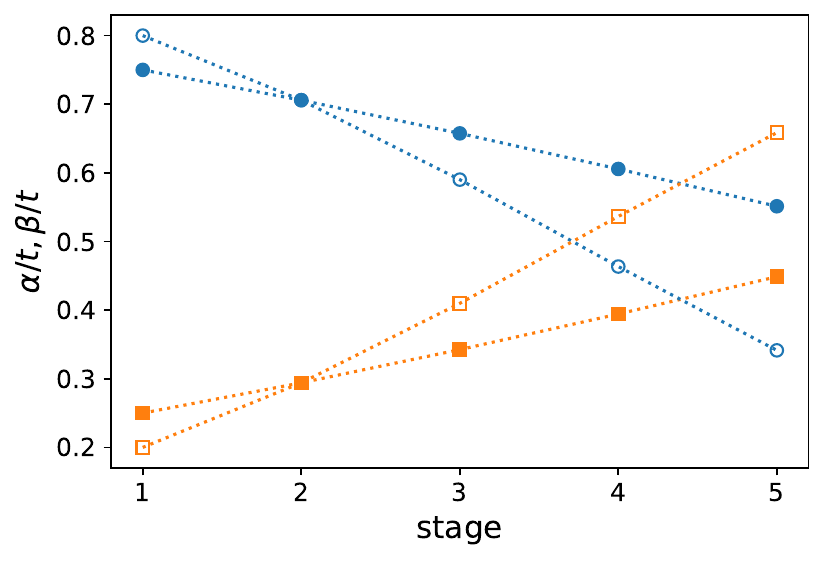}
            \caption[5 stage]%
            {5 stage}
            \label{fig:QAOA_schedule_5}
        \end{subfigure}
        \hfill
        \begin{subfigure}[b]{0.475\textwidth}  
            \centering 
            \includegraphics[width=\textwidth]{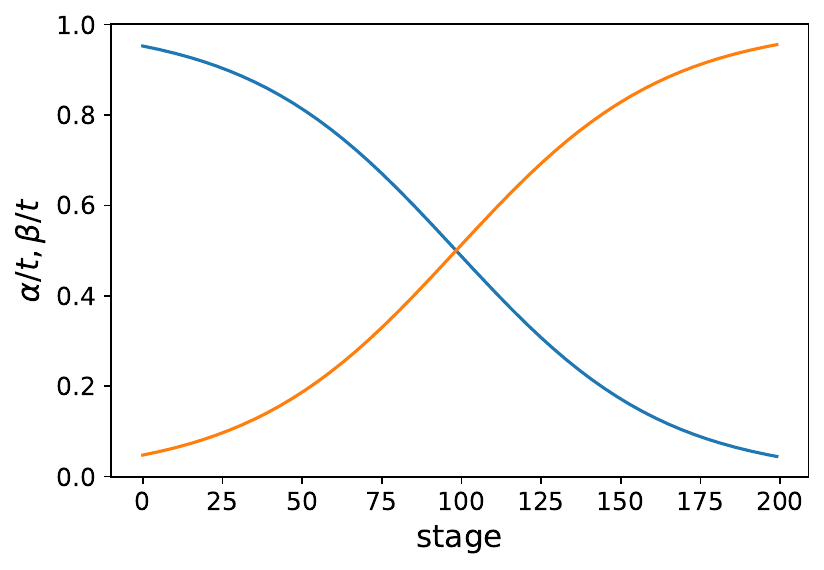}
            \caption[200 stage]%
            {200 stage}   
            \label{fig:QAOA_schedule_200}
        \end{subfigure}
        \caption[QAOA schdeules]
        {Values of $\alpha/t$ (blue, circles) and $\beta/t$ (gold, squares) for different number of stages $p$ and initial $\gamma$ values. At each stage $\gamma_{j+1}/\gamma_j= 1-\Delta\gamma$ using the parameterisation given in equation \eqref{eq:param_QAOA}. In figure (a) the filled symbols shows the approximately optimal schedule for $p=5$ with an initial $\gamma$ value of $3$ and $\Delta \gamma=0.2$ (see figure \ref{fig:5_stage_QAOA_prob}). The unfilled symbols show what it would have been for the approximately optimal values of quantum walks $\gamma=4$ and $\delta \gamma=0.2$.  Figure (b) shows that for certain parameters, $p=200$ and $\Delta \gamma=0.3$ with an initial $\gamma$ of $20$ in this case, a schedule which strongly resembles an annealing schedule can be obtained.}
        \label{fig:QAOA_schedules}
\end{figure*}

%
\begin{figure*}[htp!]
        \centering
        \begin{subfigure}[b]{0.475\textwidth}
            \centering
            \includegraphics[width=\textwidth]{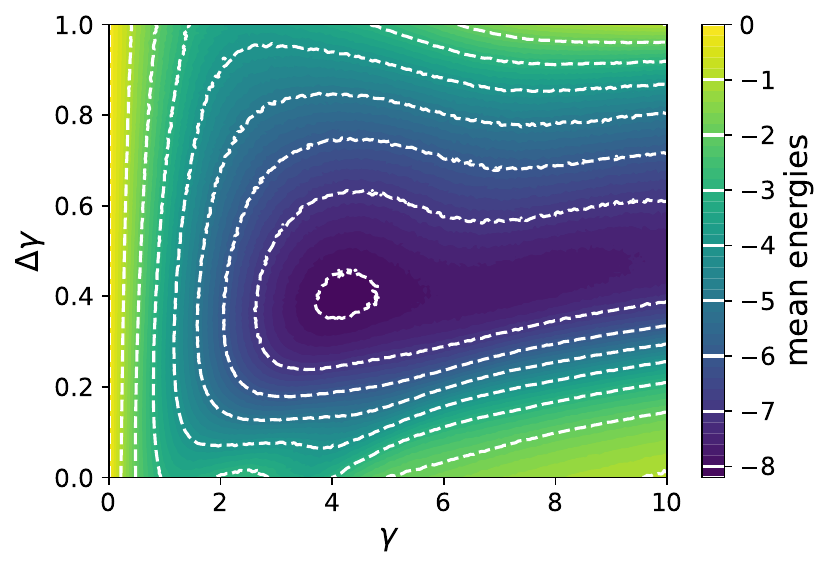}
            \caption[QW energies]%
            {QW energies}
            \label{fig:5_stage_QW_E}
        \end{subfigure}
        \hfill
        \begin{subfigure}[b]{0.475\textwidth}  
            \centering 
            \includegraphics[width=\textwidth]{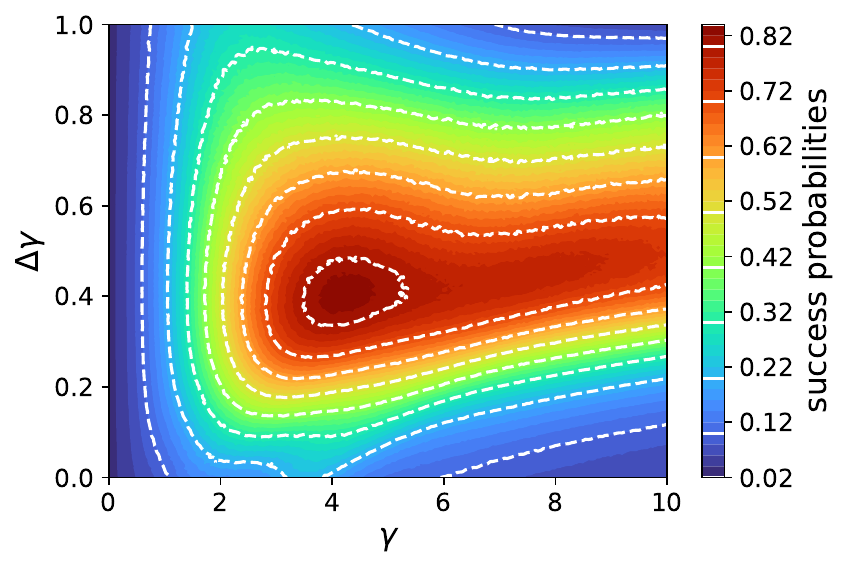}
            \caption[QW probabilities]%
            {QW probabilities}   
            \label{fig:5_stage_QW_prob}
        \end{subfigure}
        \vskip\baselineskip
        \begin{subfigure}[b]{0.475\textwidth} 
            \centering 
            \includegraphics[width=\textwidth]{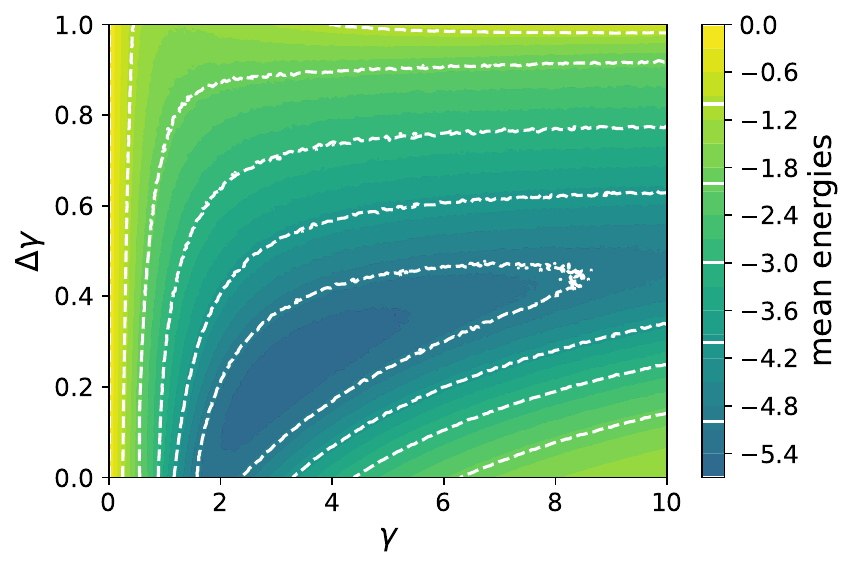}
            \caption[QAOA energies]%
            {QAOA energies}    
            \label{fig:5_stage_QAOA_E}
        \end{subfigure}
        \hfill
        \begin{subfigure}[b]{0.475\textwidth}  
            \centering 
            \includegraphics[width=\textwidth]{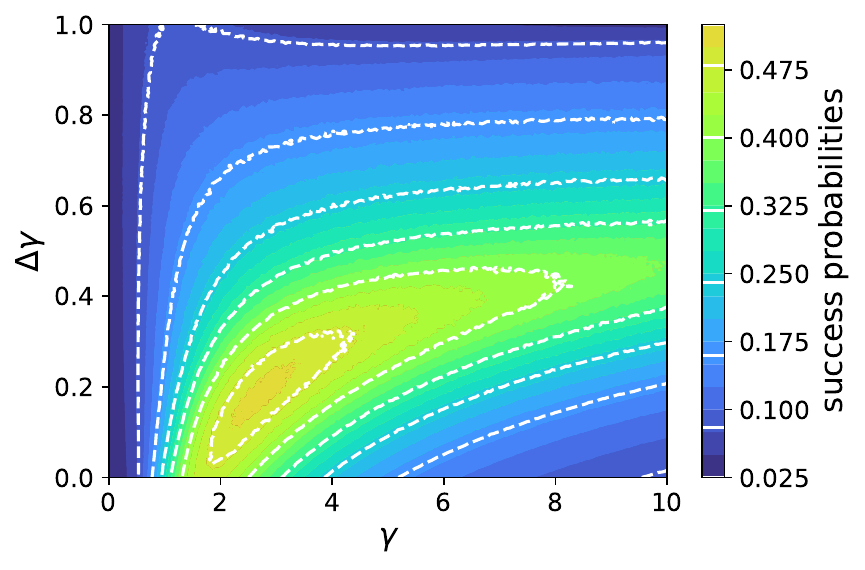}
            \caption[QAOA probabilities]%
            {QAOA probabilities}    
            \label{fig:5_stage_QAOA_prob}
        \end{subfigure}
        \caption[single_stage]
        {Quantum walk and QAOA average energies and success probabilities for five stage protocol. For quantum walk, $\gamma$ and $t$ are used to parameterise the single stage protocol. Both are parameterised in terms of $\gamma$ and $\Delta \gamma$ as described in the main text, and averaged over 2,000 samples with independent random runtimes for each stage in the range $0.1-0.5$. All plots are for for the $5$ qubit SK spin glass instance with ID: aaavmaiqiolnplcovmzxjazkyvyayz. Energy and success probability plots share the same color scale to emphasise the performance difference between the two protocols. 
        }
        \label{fig:five_stage}
\end{figure*}
%

As we consider more stages, the number of parameter increases rapidly, making the cost of classical optimization of the parameters significant. Ideally, we would like to find heuristic strategies to reduce the number of parameters, in order to reduce this extra cost, while keeping a good algorithmic performance. 

For MSQW, theoretical results \cite{Callison2021} based on energy conservation can guide our heuristic choices for $(\boldsymbol\gamma,\boldsymbol{t})$. Namely, a sequential decrease of the strength of the driver Hamiltonian $\gamma$, guarantees than the expected value of the cost function can only decrease. 
While QAOA does not meet the conditions for the energy conservation arguments in \cite{Callison2021} to apply, for small angles the conditions do hold approximately, and we confirmed this numerically.  Since optimal schedules with many stages converge to quantum annealing schedules, they do tend to follow the pattern expected for an annealing schedule, with a gradual increase of the relative contribution  of the problem Hamiltonian over the driver Hamiltonian \cite{Zhou2018qaoa}.

Inspired by these considerations, we propose the following heuristic choice of parameters for the MSQW. Since at each stage we consider the time-averaged QW, the sole parameter describing a single stage is $\gamma_j$. We choose this parameter to gradually decrease, starting from $\gamma_1=\gamma$ and evolving as $\gamma_{j+1}/\gamma_j= 1-\Delta\gamma$, for $j\in\{1, ..., p-1\}$. This induces a similar parameterization of QAOA, via equation~\eqref{eq:param_QAOA}. For reasonable choices of parameter values, these produce functions  which strongly resemble annealing schedules as shown in Fig.~\ref{fig:QAOA_schedules}. However, Fig.~\ref{fig:QAOA_schedule_5} shows that the optimal QAOA does not take on as much of an annealing-like schedule as the optimal paramerts for a multi-stage quantum walk. This is likely due to errors from the course-grained trotterization building up faster for QAOA when the problem Hamiltonian is stronger relative to the driver. 

The range of times chosen for the time-averaging was done in a similar way as discussed in section \ref{sec:two-stage}. The two-dimensional optimization landscape, described by $\gamma$ and $\Delta \gamma$ is plotted in Figs.~\ref{fig:5_stage_QW_E} and \ref{fig:5_stage_QAOA_E} for a 5-stage MSQW and QAOA (the same instance as in Fig.~\ref{fig:two_stage}), respectively, revealing a significantly superior performance of the QW approach. This also translates into a higher value of the success probability for the MSQW, which reaches values as high as 0.8 for a wide region of parameters (Fig.~\ref{fig:5_stage_QW_prob}). This is not simply an artifact of the choice of how the parameter $\gamma_j$ decreases exponentially with the stage number $j$. A different parameterization, with a linear decrease of $\Delta\gamma$, leads to the same conclusion that the QW approach outperforms QAOA, for details, see Appendix \ref{app:linear}.

\section{Discussion}\label{sec:outlook}
We have analysed the performance of multistage quantum walks as a near-term quantum computing method to solve optimization problems.  We focused on comparisons of MSQWs with QAOA. Since optimal protocols interpolating between a driver and a problem Hamiltonian usually contain smooth annealing schedules \cite{Brady2021}, we gave analytical evidence that MSQWs outperform QAOA for approximating quantum annealing, i.e., with the same number of stages we obtain a smaller error.  Via numerical simulations using instances of the spin-glass Sherrington-Kirkpatrick model as a problem Hamiltonian, we obtained further evidence that the MSQW approach outperforms QAOA. For a single stage, while the optimization landscape for QAOA has large flat regions, the QW landscape seems easier to optimize. Furthermore, over a wide region of parameters, QW is able to achieve a better performance than the best possible performance of QAOA. For multiple stages, we developed heuristic strategies for reducing the number of free parameters -- and hence the burden of classical optimization routines -- and numerically showed better performance of MSQWs over similar strategies applied to QAOA. 

Our work shows that it is promising to use the natural interactions of current quantum annealers --  where both problem and driver Hamiltonian can be applied at the same time --  to develop simple multistage heuristic protocols, and in the near term they may outperform other optimization strategies more suitable for gate-based quantum computing.  Future work analysing the performance of MSQWs for other optimization problems of interest (such as MaxCUT or Maximum Independent set), as well as a comparison to the best-known classical algorithms, would be crucial to truly understand the potential of this approach. Other options to parameterize quantum annealing functions, such as the Bezier curves used in \cite{Schulz2024guided}, can also be compared for ease of optimization and suitability for implementation in real quantum annealing hardware.

\begin{acknowledgments}
We thank Asa Hopkins for checking parts of the numerical simulations.
NC was supported by EPSRC fellowship EP/S00114X/1.
NC and VK were supported by EP/L022303/1 and impact acceleration funding associated with this grant.
VK was supported by the EPSRC UK Quantum Technology Hub in Computing and Simulation (EP/T001062/1). LN acknowledges support from FCT-Fundação para a Ciência
e a Tecnologia (Portugal) via the Project No. CEECINST/00062/2018. 
\end{acknowledgments}

\bibliography{QWQAOA}

\onecolumngrid
\appendix 

\section{Appendix: Quantum Walks as Optimal Control Protocols for Optimization Problems}
\label{app:opt}

Following the steps of the calculations done by \citeauthor{Brady2021} in \cite{Brady2021}, but with the Hamiltonian as in equation~\eqref{eq:MSQW}
we can rewrite equation~\eqref{eq:minquant2} as 
\begin{equation}
    J=\langle \psi(t_f)|\hat{H}_P|\psi(t_f)\rangle - \langle k(t_f)|\psi(t_f)\rangle+\langle k(0)|\psi(0)\rangle+\int_0^{t_f} \langle \dot{k}|\psi\rangle -i \langle k |\hat{H} |\psi\rangle\mathrm{d}t + c.c. 
\end{equation}
This is a functional on $|\psi_f\rangle, |\psi\rangle, \gamma(t)$ and $| k\rangle$, we have set the initial time to be fixed, so it does not vary. In order to get the minimum of this functional we require that $\delta J=0$ when we perturb with respect to these parameters, i.e.,
\begin{align}
0 = \delta J &= \left[ \langle \psi(t_f)|\hat{H}_P-\langle k(t_f)| \right] |\delta \psi_f\rangle + c.c.
\label{eq:FinVar}\\
&\quad\quad + \int_0^{t_f} \left[\langle\dot{k}|-i\langle k|\hat{H} \right]|\delta\psi\rangle \: \mathrm{d}t +c.c.
\label{eq:StateVar}\\
&\quad\quad + \int_0^{t_f} i\langle\psi|\frac{\partial \hat{H}}{\partial \gamma}|k\rangle \:\mathrm{d}t +c.c.
\label{eq:GamVar}\\
&\quad\quad + \int_0^{t_f} \left[-\langle\dot{\psi}|+i\langle \psi|\hat{H} \right]|\delta k\rangle\: \mathrm{d}t +c.c.
\label{eq:kvar}
\end{align}
\eqref{eq:StateVar} and \eqref{eq:kvar} imply that
\begin{align}
|\dot{\psi}\rangle &= -i\hat{H}|\psi\rangle
\label{eq:stev}\\
|\dot{k}\rangle &= -i\hat{H}|k\rangle,
\label{eq:kev}
\end{align}
which just means that the wave function and Lagrange multiplier of the ideal protocol  evolve according to the Schr\"odinger equation.\\
\eqref{eq:FinVar} gives us a condition relating the final state and final Lagrange multiplier:
\begin{equation}
|k(t_f)\rangle = \hat{H}_P |\psi(t_f)\rangle
\label{eq:finalk}
\end{equation}
Alternatively, we can rewrite equation~\eqref{eq:minquant2} as 
\begin{equation}
J = \langle \psi(0)|\hat{H}_P|\psi(0)\rangle + \int_0^{t_f} \langle \dot{\psi}|\hat{H}_P|\psi\rangle-\langle k|\dot{\psi} \rangle -i \langle k |\hat{H} |\psi\rangle +c.c. \; \mathrm{d}t.
\end{equation}
Since we do not vary the initial condition $\langle \psi(0)|\hat{H}_P|\psi(0)\rangle$ remains unchanged so we can focus on the ``action''-like part and define the Lagrangian as 
\begin{equation}
    \mathcal{L} = \langle \dot{\psi}|\hat{H}_P|\psi\rangle -\langle k|\dot{\psi} \rangle -i \langle k |\hat{H} |\psi\rangle +c.c.
\end{equation}
Then we can define the ``momenta''
\begin{align*}
\langle p| &= \frac{\partial \mathcal{L}}{\partial (|\dot{\psi}\rangle)} = \langle\psi|\hat{H}_P - \langle k| \quad \text{and} \\
| p\rangle &=\frac{\partial \mathcal{L}}{\partial (\langle\dot{\psi}|)} = \hat{H}_P|\psi\rangle - |k\rangle,
\end{align*}
which we use to construct the control Hamiltonian
\begin{equation}
\mathcal{H} = \langle p|\dot{\psi}\rangle + \langle\dot{\psi}|p\rangle - \mathcal{L} = i\langle k|\hat{H}|\psi\rangle - i\langle\psi|\hat{H}| k\rangle
\end{equation}
Since the Lagrangian has no explicit time dependence the Hamiltonian should not change with time, i.e., $\frac{\mathrm{d}\mathcal{H}}{\mathrm{d}t}=0$, and so
\begin{equation}
0 = \frac{\mathrm{d}\mathcal{H}}{\mathrm{d}t} = i\langle k|\frac{\partial\hat{H}}{\partial t}|\psi\rangle - i\langle\psi|\frac{\partial\hat{H}}{\partial t}| k\rangle.
\end{equation}
Since $\frac{\partial\hat{H}}{\partial t}=\dot{\gamma}\hat{H}_d$ we have
\begin{equation}
0 = i\langle k|\dot{\gamma}\hat{H}|\psi\rangle - i\langle\psi|\dot{\gamma}\hat{H}_d| k\rangle \equiv \dot{\gamma}\Phi_d(t),
\label{eq:apptime}
\end{equation}
where we define $\Phi_X(t)=i\langle k|\hat{H}_X|\psi\rangle+c.c.$. We can then rewrite $\mathcal{H}$ as 
\begin{equation}
\mathcal{H} = \gamma(t)\Phi_d(t)+\Phi_P(t).
\label{eq:appcontham}
\end{equation}
Making use of equation~\eqref{eq:finalk} we get
\begin{equation}
    \Phi_P(t_f) = i\langle k(t_f)|\hat{H}_P|\psi(t_f)\rangle+c.c.
    =i\langle \psi(t_f)|\hat{H}_P^2|\psi(t_f)\rangle+c.c. = 0.
\end{equation}
And hence $\mathcal{H} = \gamma(t_f)\Phi_d(t_f)$.\\
Similarly, since we start the system in the ground state of $\hat{H}_d$ we can set $\hat{H}_d|\Psi(0)\rangle=0$. (Note that if this is not already the case by the construction of $\hat{H}_d$ we can add or subtract a multiple of the identity operator to adjust it without changing the dynamics.) Then
\begin{equation*}
\Phi_d(0) =  i\langle k(0)|\hat{H}_d|\Psi(0)\rangle+c.c.= 0.\\
\end{equation*}
And so $\mathcal{H}=\Phi_P(0)$.\\
Differentiating equation~\eqref{eq:appcontham} with respect to time gives us
\begin{equation}
    \frac{\mathrm{d}\mathcal{H}}{\mathrm{d}t} = \dot{\gamma}(t)\Phi_d(t) + \gamma(t)\dot{\Phi}_d(t) + \Phi_P(t).
\end{equation}
From equation~\eqref{eq:apptime} $\frac{\mathrm{d}\mathcal{H}}{\mathrm{d}t}=\dot{\gamma}\Phi_d(t)=0$, so 
\begin{equation}
    0 = \dot{\gamma}(t)\Phi_d(t) + \gamma(t)\dot{\Phi}_d(t) + \Phi_P(t) = \gamma(t)\dot{\Phi}_d(t) + \Phi_P(t)
      \;\iff\;    \dot{\Phi}_P(t) = -\gamma(t)\dot{\Phi}_d(t).
\end{equation}

\section{Appendix: Approximation of quantum annealing with quantum walks and QAOA}
\label{app:approx}

The total time evolution for the annealing schedule is given by
\begin{equation}
    \hat{U}_{QA}(t_0,t_f) = \mathcal{T}\left[e^{-i\int_{t_{0}}^{t_f}\hat{H}(t) \,\mathrm{d}t}\right],
\end{equation}
The Hamiltonian is the annealing Hamiltonian $\hat{H}(t)=A(t)\hat{H}_d+B(t)\hat{H}_P$\\
We split the total time into multiple small segments $(t_{j-1},t_j)$, which satisfy $\|\hat{H}\|_{max}(t_j-t_{j-1})<1$ and expand the evolution in each segment into its Dyson series
\begin{equation}
    \hat{U}_{QA} = \mathcal{T}\left[e^{-i\int_{t_{j-1}}^{t_j}\hat{H}(t) \,\mathrm{d}t}\right]= \mathbb{1} -i\int_{t_{j-1}}^{t_j}\hat{H}(t) \,\mathrm{d}t- \int_{t_{j-1}}^{t_j}\int_{t_{j-1}}^t\hat{H}(t)\hat{H}(t') \,\mathrm{d}t'\,\mathrm{d}t+ ...\\
\label{eq:AppDyson}
\end{equation}
Now defining $
    \alpha_j = \frac{1}{t_j-t_{j-1}}\int_{t_{j-1}}^{t_j} A(t) \:\mathrm{d}t$, $
    \beta_j = \frac{1}{t_j-t_{j-1}}\int_{t_{j-1}}^{t_j} B(t) \:\mathrm{d}t$ 
and $\tilde{H} = \alpha_j\hat{H}_d+\beta_j\hat{H}_P$ from Taylor's Theorem we get the approximation
\begin{equation}
    H(t) = \tilde{H} +\mathcal{O}\left(\dot{H}_{max}(t_j-t_{j-1})\right) \Rightarrow \int_{t_{j-1}}^{t} H(t) \:\mathrm{d}t = \tilde{H}(t-t_{j-1}) + \mathcal{O}\left(\dot{H}_{max}(t_j-t_{j-1})^2\right),
    \label{eq:TayApprox}
\end{equation}
where we define $\dot{H}_{max}=\max\limits_{t\in(t_0,t_f)}\|\dot{H}(t)\|$.\\

Substituting these approximations into \eqref{eq:AppDyson}
\begin{align}
\begin{split}
\hat{U}_{QA}&= 
 \mathbb{1} -i\tilde{H}(t_j-t_{j-1}) -
  \int_{t_{j-1}}^{t_j} \left[\tilde{H} +\mathcal{O}\left(\dot{H}_{max}(t_j-t_{j-1})\right)\right]\\
  &\quad\quad\quad\quad\quad\quad\quad\quad\quad\quad\quad\quad\times\left[\tilde{H}(t-t_{j-1}) + \mathcal{O}\left(\dot{H}_{max}(t_j-t_{j-1})^2\right)\right]\mathrm{d}t + ...\\
&= \mathbb{1} -i\tilde{H}(t_j-t_{j-1}) -
  \int_{t_{j-1}}^{t_j} \tilde{H}^2(t-t_{j-1}) + \mathcal{O}\left(\|\tilde{H}\| \dot{H}_{max}(t_j-t_{j-1})^2\right) \mathrm{d}t + ...\\
  &=\mathbb{1} -i\tilde{H}(t_j-t_{j-1}) -
  \frac{1}{2} \tilde{H}^2(t_j-t_{j-1})^2 + ... +\mathcal{O}\left(H_{max} \dot{H}_{max}(t_j-t_{j-1})^3\right).
\end{split}
\label{eq:QAApprox}
\end{align}
Expanding the time evolution operator for a QAOA step with the same parameters $\alpha_j$ and $\beta_j$ gives us
\begin{align*}
\hat{U}_{QAOA} &=  e^{-i\alpha_j\hat{H}_d(t_j-t_{j-1})}e^{-i\beta_j\hat{H}_P(t_j-t_{j-1})}\\
&= \mathbb{1} -i\tilde{H}(t_j-t_{j-1}) -
  \frac{1}{2} \tilde{H}^2(t_j-t_{j-1})^2 - \frac{1}{2}\left[\hat{H}_d,\hat{H}_P\right]\alpha_j\beta_j(t_j-t_{j-1})^2 + \mathcal{O}\left(\|\tilde{H}\|^3(t_j-t_{j-1})^3\right).
\end{align*}
The difference between the annealing and QAOA evolutions is thus
\begin{equation}
\|\hat{U}_{QA} - \hat{U}_{QAOA}\| = \frac{1}{2}\left\|\left[\hat{H}_d,\hat{H}_P\right]\right\|\alpha_j\beta_j(t_j-t_{j-1})^2 +  \mathcal{O}\left(H_{max}^3(t_j-t_{j-1})^3 + H_{max} \dot{H}_{max}(t_j-t_{j-1})^3\right)  \end{equation}
Expanding the multi-stage quantum walk dynamics for a single stage gives us
\begin{align*}
\hat{U}_{QW} &=  e^{-i(\alpha_j\hat{H}_d+\beta_j\hat{H}_P)(t_j-t_{j-1})}= \mathbb{1} -i\tilde{H}(t_j-t_{j-1}) -
  \frac{1}{2} \tilde{H}^2(t_j-t_{j-1})^2 +...
\end{align*}
So we have
\begin{equation}
    \hat{U}_{QA} - \hat{U}_{QW} = \mathcal{O}\left(H_{max} \dot{H}_{max}(t_j-t_{j-1})^3\right)
\end{equation}
Now we take each $t_j-t_{j-1}=\frac{t_f-t_0}{p}$, i.e. all the segments have equal length. Since the total dynamics consist of all the segments put together the errors in each stage add up, giving us
\begin{align*}
\|\hat{U}_{QA}(t_0,t_f) - \hat{U}_{QAOA}(t_0,t_f)\| &= \frac{1}{2}\left\|\left[\hat{H}_d,\hat{H}_P\right]\right\|\left(\frac{t_f-t_0}{p}\right)^2\sum_{j=1}^p\alpha_j\beta_j \\
&\quad+\mathcal{O}\left(\|H\|_{max}^3\frac{(t_f-t_0)^3}{p^2} + \|H\|_{max} \|\dot{H}\|_{max}\frac{(t_f-t_0)^3}{p^2}\right)  
\end{align*}
The sum term scales linearly with $p$, so in total the error scales as $\frac{1}{p}$.\\
For the MSQW
\begin{equation}
    \|\hat{U}_{QA}(t_0,t_f)-\hat{U}_{QW}(t_0,t_f)\| = \mathcal{O}\left(H_{max} \dot{H}_{max}\frac{(t_f-t_0)^3}{p^2}\right) 
\end{equation}
So the error here scales as $\frac{1}{p^2}$. 
Hence the MSQW not only approximates Quantum Annealing better than QAOA since it doesn't come with the extra commutator term, it also approaches it faster as we increase the number of steps.\\
For the second order Trotterization the error for each step \cite{Childs2021_TrotterError} is 
\begin{equation*}
    \|U_{QW}-U_{QAOA}\| \leq \frac{(t_j-t_{j-1})^3}{12}\|\left[H_P,\left[H_P,H_d\right]\right]\|+\frac{(t_j-t_{j-1})^3}{24}\|\left[H_d,\left[H_d,H_P\right]\right]\| = \mathcal{O}\left(H_{max}^3(t_j-t_{j-1})^3\right),
\end{equation*}
which results in a cumulative error

\begin{align*}
    \left\| \hat{U}_{QA}(t_0,t_f)-\hat{U}_{QAOA}(t_0,t_f) \right\| &\leq \left\| \hat{U}_{QA}(t_0,t_f)-\hat{U}_{QW}\right\| + \left\| \hat{U}_{QW}(t_0,t_f)-\hat{U}_{QAOA}(t_0,t_f)\right\|\\
    &= \mathcal{O}\left(H_{max}^3\frac{(t_f-t_0)^3}{p^2} + H_{max} \dot{H}_{max}\frac{(t_f-t_0)^3}{p^2}\right) ,
\end{align*}
which again scales like $\frac{1}{p^2}$, although there is still an additional term which the MSQW expansion does not have.

\section{Appendix: Linear Parameterization \label{app:linear}}

%
\begin{figure*}[htp!]
        \centering
        \begin{subfigure}[b]{0.475\textwidth}
            \centering
            \includegraphics[width=\textwidth]{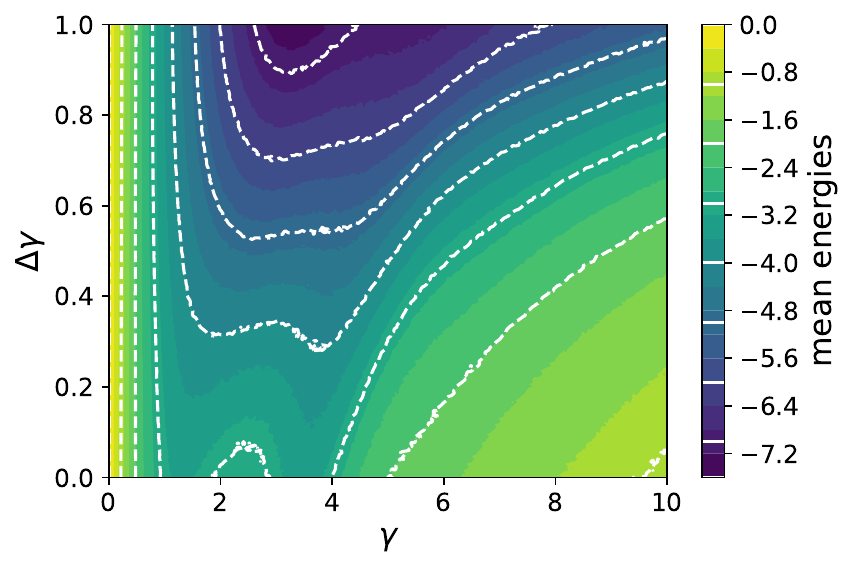}
            \caption[QW energies]%
            {QW energies}
            \label{fig:5_stage_QW_E_lin}
        \end{subfigure}
        \hfill
        \begin{subfigure}[b]{0.475\textwidth}  
            \centering 
            \includegraphics[width=\textwidth]{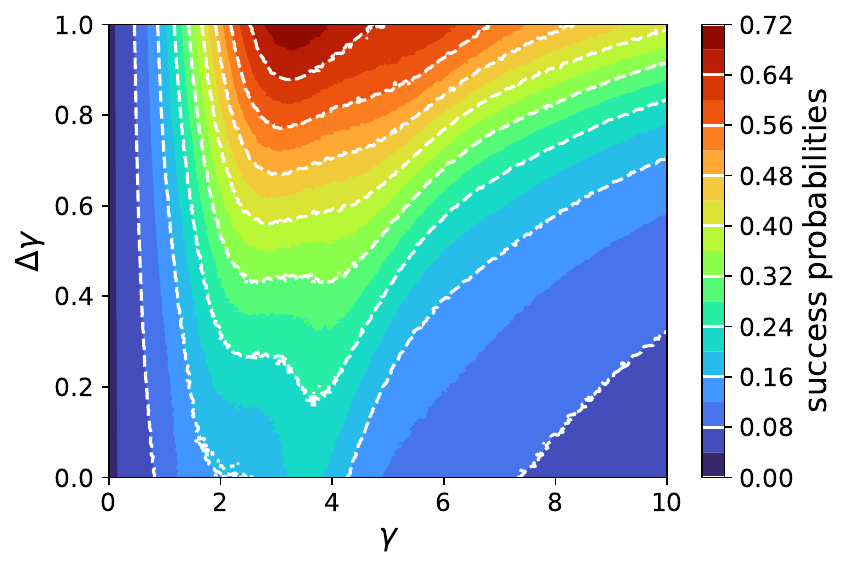}
            \caption[QW probabilities]%
            {QW probabilities}   
            \label{fig:5_stage_QW_prob_lin}
        \end{subfigure}
        \vskip\baselineskip
        \begin{subfigure}[b]{0.475\textwidth} 
            \centering 
            \includegraphics[width=\textwidth]{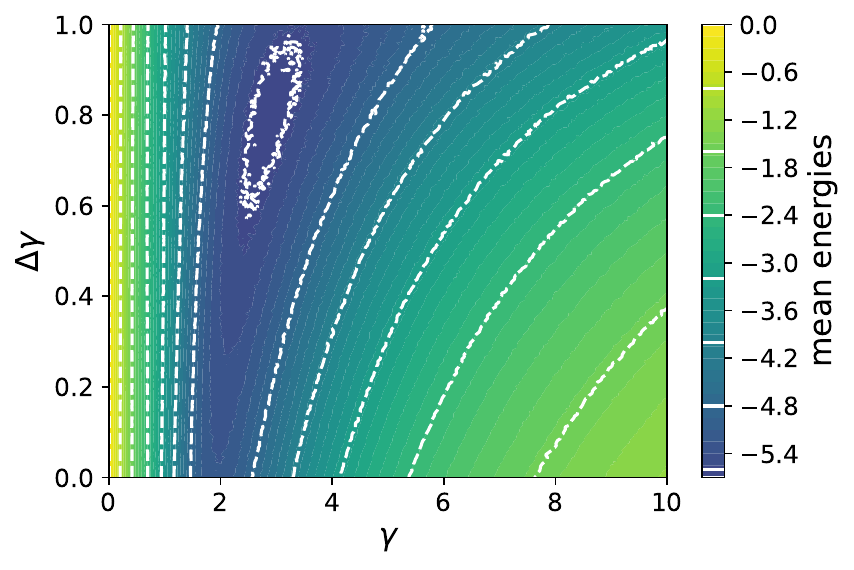}
            \caption[QAOA energies]%
            {QAOA energies}    
            \label{fig:5_stage_QAOA_E_lin}
        \end{subfigure}
        \hfill
        \begin{subfigure}[b]{0.475\textwidth}  
            \centering 
            \includegraphics[width=\textwidth]{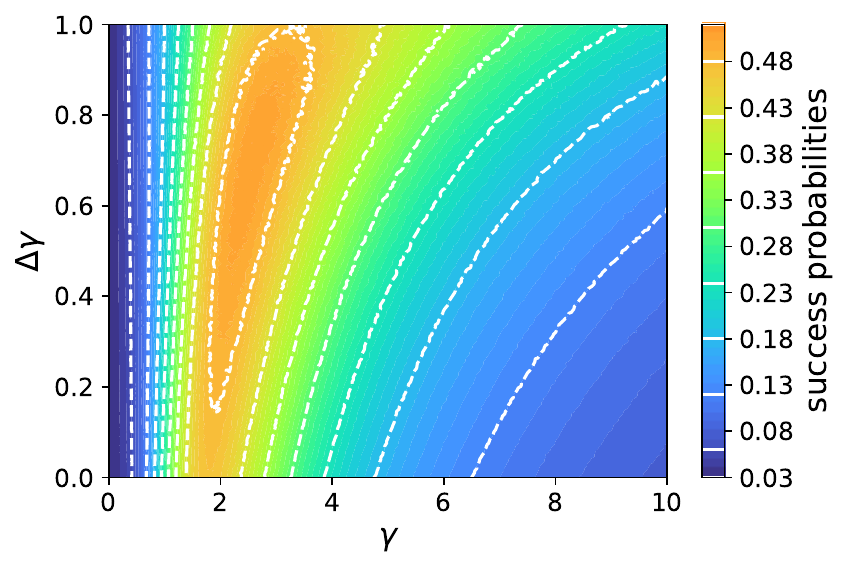}
            \caption[QAOA probabilities]%
            {QAOA probabilities}    
            \label{fig:5_stage_QAOA_prob_lin}
        \end{subfigure}
        \caption[single_stage]
        {Quantum walk and QAOA average energies and success probabilities for five stage protocol, but with a linear sweep in $\gamma$. For quantum walk, $\gamma$ and $t$ are used to parameterise the single stage protocol. Both are parameterised in terms of $\gamma$ and $\Delta \gamma$ as described in the main text, and averaged over 2,000 samples with independent random runtimes for each stage in the range $0.1-0.5$. All plots are for for the $5$ qubit SK spin glass instance with ID: aaavmaiqiolnplcovmzxjazkyvyayz. Energy and success probability plots share the same color scale to emphasise the performance difference between the two protocols. This is the version with a linear schedule. 
        }
        \label{fig:five_stage_linear}
\end{figure*}

Figure \ref{fig:five_stage_linear} depicts the results of a five stage quantum walk versus QAOA with a protocol which is linear meaning at stage $k$ (with $k$ starting from $0$ and running to $4$): $\gamma_k=\gamma-k\,\frac{\Delta \gamma}{\gamma}$. While the shape of the plots are slightly different the results are qualitatively the same, quantum walk significantly out-performs QAOA.

\end{document}